\documentclass[aps, prb, reprint, longbibliography]{revtex4-2}
\usepackage{amsmath,amsfonts,amssymb}
\usepackage{graphicx}
\usepackage[colorlinks=true, allcolors=blue]{hyperref}
\usepackage{soul}
\usepackage{array}
\usepackage[table,xcdraw]{xcolor}

\newcolumntype{P}[1]{>{\centering\arraybackslash}p{#1}}

\usepackage{hyperref} 

\begin{document} 

\title{
Josephson tunnel junction arrays and Andreev weak links: linked by a single energy-phase relation
}

\author{A. Mert Bozkurt}
\email{a.mertbozkurt@gmail.com}
\affiliation{Kavli Institute of Nanoscience and QuTech, Delft University of Technology, P.O. Box 4056, 2600 GA Delft, The Netherlands}
\author{Valla Fatemi}
\email{vf82@cornell.edu}
\affiliation{School of Applied and Engineering Physics, Cornell University, Ithaca, NY, 14853, USA}

\date{\today}

\begin{abstract}

Josephson elements are cornerstones of cryogenic classical and quantum superconducting technology, owing to their nonlinearity.
Two important types of Josephson elements are often considered distinct: the tunnel junction (superconductor-insulator-superconductor, SIS) and the Andreev weak link (superconductor-normal-superconductor, SNS) referring to any non-superconducting and non-insulating central region.
SNS junctions and SIS junctions have appeared in related technological and fundamental science contexts over the last decade, such as in the design of protected qubit concepts. 
In this perspective article, we review correspondences between SISIS junctions and SNS junctions in limiting regimes, in which a single energy-phase relationship describes both systems. 
We show how this insight helps to connect recent bodies of theoretical and experimental work in both systems, and conclude by describing a few important differences.
\end{abstract}
\maketitle

\onecolumngrid

\section{Introduction}
\label{sec:intro}
A single, conventional Josephson tunnel junction exhibits a tantalizingly simple, cosinusoidal energy-phase relation~\cite{josephson_possible_1962}.
Although itself powerful in many situations, higher harmonics in the energy-phase relation can be useful and produce intriguing phenomena.
Such higher harmonics can arise when superconducting reservoirs are hooked up to a central region that hosts an internal degree of freedom. 
The two most prominent examples of this, shown in Fig.~\ref{fig:theone}(a-b), are superconductor-normal-superconductor junctions (SNS)\footnote{Here, we use ``normal'' to refer to any non-superconducting and non-insulating central region, not just metals with high electron density. Additionally, note that we do not consider strongly correlated states in the normal region like magnetism.} and arrays of superconductor-insulator-superconductor (SIS) junctions.
SNS junctions host microscopic Andreev bound states inside the junction, while the SISIS array (the minimal array) hosts superpositions of charge states of the central island. 
Here, we will collect connections, both physical and mathematical, between these two cases. 
We will then show how such connections have been developed in parallel, but based on the same basic physics, to accomplish certain applied or basic device concepts.
A common (though not universal) connective element is the idea of engineering the harmonic content (the Fourier components of the energy phase relation) of a superconducting structure. 
We conclude by discussing some important differences that motivate their continued separate development. 

We proceed with two caveats: we assume some familiarity with superconductivity~\cite{tinkham_introduction_2004}, mesoscopic electronics~\cite{hanson_spins_2007,prada_andreev_2020}, and Josephson physics in circuits~\cite{vool_introduction_2017}, and we apologize in advance that the references may not be comprehensive. 

\section{The energy-phase relationship}
\label{sec:epr}
\begin{figure}[htpb]
\centering
\includegraphics[width=\linewidth]{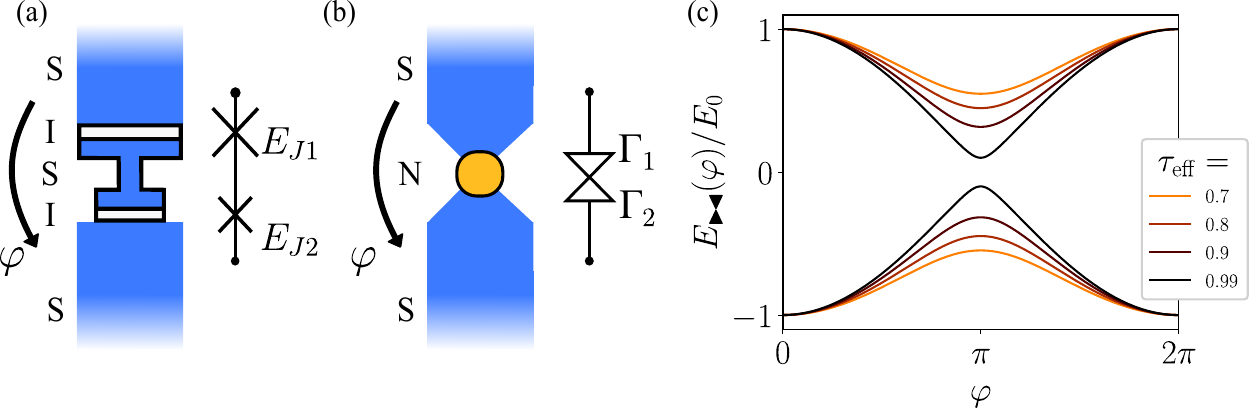}
\caption{
Schematics of a two terminal structure with outer superconducting reservoirs and a central region with (a) a metallic superconductor coupled to the reservoirs with insulating tunnel barriers and 
(b) a normal (non-superconducting and non-insulating) region which is coupled to the reservoirs with Andreev reflection. 
For the normal case, we consider in particular a short normal region, such as a resonant level (as may be hosted in a quantum dot) or a quantum point contact region.
For each setup, the circuit symbols are shown next to the schematic. 
(c) EPR in units of $E_0$ for a single arm with various values of $\tau_\mathrm{eff}$. 
For SNS and weak-coupling SISIS, these are quantum levels. For strong-coupling SISIS, these are the classically allowed solutions (see text and Table~\ref{table:1}).}
\label{fig:theone}
\end{figure}

At the center of our discussion is the following energy-phase relation (EPR):
\begin{equation}
E(\varphi) = \pm E_0 \sqrt{1 - \tau_\mathrm{eff} \sin^2(\varphi/2)}, \label{eq:theone}
\end{equation}
where $\varphi$ is the superconducting phase difference across the junction in question, $E_0$ is an energy scale, and $0\leq \tau_\mathrm{eff}\leq1$ is an effective transparency.
The parameter $\tau_\mathrm{eff}$ controls the harmonic content of the EPR: for $\tau_\mathrm{eff}\ll1$ a conventional cosine Josephson relation is recovered, $E \approx \pm(E_0 + E_0 \tau_\mathrm{eff} \cos(\varphi) /4)$, whereas higher harmonics are prominent for $ 1 - \tau_\mathrm{eff} \ll 1$ (see Fig.~\ref{fig:theone}(c)).
These harmonics are crucial for designing EPRs tailored to different applications (more on this later).
As we will see, the EPR~\eqref{eq:theone} shows up for both  SNS and SISIS junctions in two extreme limits of coupling, which refers to how strongly the central region (N or S) is connected to the outer superconducting reservoirs.
Finally, the $\pm$ sign in Eq.~\eqref{eq:theone} is indicative of an internal degree of freedom of the central region.

\begin{table}[htbp]
\centering
\large
\begin{tabular}[c]{ | P{5.5em} | P{3.3cm}| P{3.3cm} | P{3.3cm} | P{3.3cm} | }
\cline{2-5}
\multicolumn{1}{c|}{} & \multicolumn{2}{|c|}{SNS} & \multicolumn{2}{|c|}{SISIS} \\\cline{2-5}
\multicolumn{1}{c|}{} & \multicolumn{1}{|c|}{\normalsize weak coupling} & \multicolumn{1}{|c|}{\normalsize strong coupling} & \multicolumn{1}{|c|}{\normalsize weak coupling} & \multicolumn{1}{|c|}{\normalsize strong coupling} \\\cline{2-5}
\hline
$E_0$ & $\sqrt{\Gamma_\Sigma^2 + \epsilon_{\textrm{QD}}^2}$ & $\Delta$ & \normalsize$\sqrt{E_{J\Sigma}^2 + E_C^2(1-n_g)^2}$  & $E_{J\Sigma}$ \\ 
 \hline
 \Large$\tau_{\textrm{eff}}$ & \Large$\frac{4\Gamma_1\Gamma_2}{\Gamma_\Sigma^2 + \epsilon_{\textrm{QD}}^2}$ & \Large$\tau$ & \Large$\frac{4E_{J1}E_{J2}}{E_{J\Sigma}^2 + E_C^2(1-n_g)^2}$ & \Large$\frac{4E_{J1}E_{J2}}{E_{J\Sigma}^2}$\\
 \hline
 \normalsize
 Internal degree of freedom, providing $\pm$ &  \normalsize
Quantum superposition of $N$ and $N+2$ electrons &  \normalsize
Quantum superposition of many electron configurations & \normalsize Quantum superposition of $N$ and $N+2$ electrons &  \normalsize
Classical mean phase $\varphi_1(\varphi) + \{0, \pi\}$ \\
 \hline
\end{tabular}
\vskip 1 em
\caption{Table that summarizes the mathematical and physical correspondences. Here, we define $\Gamma_\Sigma \equiv \Gamma_1 + \Gamma_2$, $E_{J\Sigma}\equiv E_{J1} + E_{J2}$.}
\label{table:1}
\end{table}

\subsection{Weak coupling limit} 
Perhaps the more well-known correspondence between the SNS and SISIS emerges in the regime where the central region is weakly coupled to the reservoirs.
For SNS the simplest case is that of a resonant level: a small quantum dot with level spacing much larger than the superconducting gap $\Delta$ that is weakly tunnel-coupled to the reservoirs (S-QD-S)~\cite{beenakker1992Resonant, wendin1996Josephson,kurilovich_microwave_2021}.
For SISIS this regime is called the Cooper-pair transistor (CPT), in which the metallic superconducting island has charging energy $E_C \gg E_J$, where $E_J$ is the Josephson energy scale~\cite{joyez_single_1995}. 
In both cases, one isolates two charge states as a basis: $N$ and $N+2$ electrons on the central region (here we focus on the even parity sector that is near resonance~\footnote{Even-odd fermion parity transitions are a subject of study for both CPT~\cite{joyez_single_1995,aumentado2004Nonequilibrium} and S-QD-S~\cite{martin-rodero2011Josephson, kadlecova_practical_2019,kurilovich_microwave_2021,bargerbos_singlet-doublet_2022,fatemi_microwave_2022}, in part due to quasiparticle poisoning~\cite{glazman_bogoliubov_2021}. Unlike S-QD-S, the CPT features an inherent energy penalty $\Delta$ for odd parity states, making them less energetically favorable compared to even parity states. However, we note that, depending on specific parameter values, such as $\epsilon$, $\gamma_i$, and $\varphi_i$, and charging effects, the opposite charge parity state $(N + 1)$ can become energetically favorable in both systems.}).
Then, these two systems can have identical low-energy Hamiltonians, up to the name and meaning of the parameters:
\begin{equation}
H = \epsilon \sigma_z +  \left( \gamma_1 e^{i\varphi_1} + \gamma_2 e^{i\varphi_2} \right) \sigma_+ + \textrm{h.c.},\label{eq:cpt_hamiltonian}
\end{equation}
with $\sigma_z$ denoting Pauli matrix and $\sigma_{\pm}$ denoting the raising and lowering operators in the charge state basis, and $\gamma_i$ representing the coupling strength to reservoir $i$.
Here, $\varphi_{1,2}$ denote the phases of the superconducting reservoirs with $\varphi = \varphi_1 - \varphi_2$.
The couplings can be the tunnel couplings $\gamma_i = \Gamma_i\ll\Delta$ in S-QD-S ($\Delta$ is the superconducting gap)~\cite{beenakker_resonant_1992} or the Josephson energies $\gamma_i = E_{J,i}\ll E_C$ of each tunnel junction in CPT~\cite{friedman2002AharonovCasherEffect}. 
In the same fashion, $\epsilon$ can be the energy level offset $\epsilon_{\textrm{QD}}$ for S-QD-S and $\epsilon = E_C (1-n_g)$ for the CPT, where $n_g$ is the charge offset of the island (in units of $2e$). 
The eigenenergies of this Hamiltonian have the form Eq.~\eqref{eq:theone}, with $E_0 \equiv \sqrt{(\gamma_1 + \gamma_2)^2 + \epsilon^2}$ and $\tau_\mathrm{eff} \equiv 4\gamma_1\gamma_2 / E_0^2$.
Thus, within this manifold of states, these two physically distinct systems behave essentially identically in the weak coupling limit, both in terms of their internal charge degree of freedom, described by Pauli matrices $\mathbf{\sigma}$, and the EPR seen by a surrounding circuit.

\subsection{Strong coupling limit} 
In the limit of strong coupling between the central region and the reservoirs, an unexpected correspondence emerges that is distinct from the full correspondence observed in the weak coupling regime.
A short SNS junction with $\Gamma_i \gg \Delta$ retains discrete quantum levels, Andreev bound states (ABS), which disperse as per Eq.~\eqref{eq:theone} with $E_0 = \Delta$ and $\tau_n$ for each conduction channel $n$~\cite{beenakker_universal_1991}, which have been observed in, e.g., aluminum atomic point contacts~\cite{bretheau_localized_2013}.
A single-channel, short SNS junction retains a single pair of ABS, which can be manipulated as a qubit degree of freedom~\cite{janvier_coherent_2015}.

On the other hand, the central region of the SISIS junction should exhibit classical behavior in this limit, $E_{J,i}\gg E_C$.
In this limit, the phase of the island is a classical quantity and quantum fluctuations can be ignored.
Although this might suggest no relationship between the two device types, it turns out that the SISIS EPR also follows Eq.~\eqref{eq:theone}~\cite{bozkurt2023DoubleFourier}. 
This is found by applying current conservation,~\cite{bozkurt2023DoubleFourier, frattini2018Optimizing}
\begin{equation}
E_{J1}\sin(\varphi_1) = E_{J2}\sin(\varphi - \varphi_1), \label{eq:current_conservation}
\end{equation}
where $\varphi_1$ is the phase difference across only the first tunnel junction.
For a given $\varphi$, current conservation constrains $\varphi_1$ to two possible values differing by $\pi$, which serves as a classical internal degree of freedom.
Thereby, we recover Eq.~\eqref{eq:theone}, with $E_0 \equiv E_{J1} + E_{J2} $ and $\tau_{\mathrm{eff}} \equiv 4 E_{J1} E_{J2} / E_0^2$~\footnote{An important caveat is that for exactly $\tau_\mathrm{eff}=1$ and $\varphi=\pi$, the Josephson terms are frustrated and the island cannot be treated classically.}. 
We remark that, in this case, we arrived at Eq.~\eqref{eq:theone} by applying a constraint (current conservation) to a scalar function (the total energy), whereas in the other cases Eq.~\eqref{eq:theone} is arrived at through matrix eigenvalues. 
In practice, the strong-coupling approximation holds when the quantum fluctuations of the central island phase are small while also keeping the frequency of the associated collective mode of the island well above the dynamics of $\varphi$~\cite{rymarz2023consistent}, which can be made slow by a capacitive shunt.
This type of design is frequently used in circuit quantum electrodynamics contexts~\cite{manucharyan_superinductance_2012,frattini_three-wave_2021}.

\section{What insights can we draw from these correspondences? }
One conclusion is that if the dynamics of the internal degrees of freedom (whether the internal degree of freedom is an ABS or charge on the metallic island) are irrelevant to the context at hand, the behavior of SISIS and short, single-channel SNS junctions, at both the strong- and weak-coupling limits, is the same. 
This is because they all share the ground state EPR given by the low-energy branch of Eq.~\eqref{eq:theone}.
Further, the S-QD-S and resonant level Hamiltonians behave the same even if one includes the higher energy branch.
These observations inspired investigations to check if proposals or experimental observations in SNS junctions can be also be realized in SIS junction arrays, which are technologically more mature.
Alternatively, concepts for SIS arrays can take advantage of the experimental features of SNS junctions, such as their gate tunability (see Sec.~\ref{sec:discussion}).
In the following, we consider some examples from the literature to showcase the correspondence between the two types of devices.

\begin{figure}[htbp]
\centering
\includegraphics[width=0.8\linewidth]{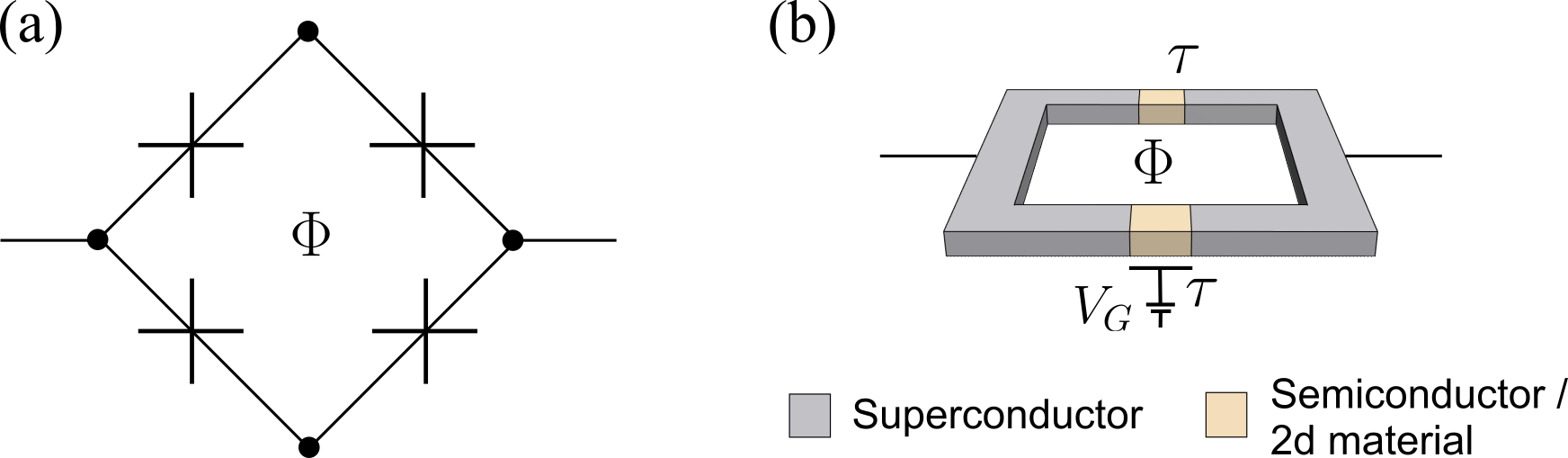}
\caption{Single-loop interferometers of (a) the SISIS array, known as the rhombus and (b) SNS junctions. 
In both circuits, at half flux $\Phi=\Phi_0/2$, where $\Phi_0=h/2e$ is the superconducting flux quantum, the parity of transferred Cooper pairs become decoupled manifolds.
When the arms are imbalanced, such interferometers can also be used to create superconducting diodes.
Additionally, SNS junction properties can be tuned by applying a gate voltage $V_G$.
Panel (b) was adapted from Schrade \& Fatemi~\cite{schrade2024Dissipationless}.
}
\label{fig:protected_qubit}
\end{figure}

\textit{Superconducting Diode Effect:} The superconducting diode effect refers to unequal critical currents in opposite directions.
Diode effect in interferometers with imbalanced loop-inductance was observed shortly after the Josephson effect itself~\cite{goldman_meissner_1967,fulton_quantum_1972,likharev_dynamics_1986}.
Recently, there have been proposals and experiments on realizing superconducting diode effect using single-loop interferometers with high-transparency SNS junctions, aiming particularly to take advantage of the gate-tunability~\cite{fominov2022Asymmetric, souto2022Josephson, ciaccia2023Gate, valentini2023Radio}.
By controlling the magnetic flux piercing the interferometer loop and the individual transparency  $\tau_\mathrm{eff}$ of each arm, a superconducting diode effect can be achieved.
Drawing on our recent observation that classical SISIS junctions exhibit the EPR of Eq.~\eqref{eq:theone}, we can bring these two approaches into direct correspondence. 
The fact that one parameter, $\tau_\mathrm{eff}$, controls the harmonic content of each SISIS arm makes systematic engineering of complicated EPRs more tractable, including optimization of the diode effect~\cite{bozkurt2023DoubleFourier}.
By choosing the Josephson energies of each tunnel junction and controlling the flux passing through the array, it is possible to design the overall EPR of the array to accommodate various EPRs, including ones that result in a superconducting diode.

\textit{Cooper-pair-parity Protected Qubits:}
The same kind of interferometers as in the diodes above have been considered for qubits that are protected by conservation of Cooper pair parity, both in the SISIS case (known as the rhombus)~\cite{gladchenko_superconducting_2009,bell_protected_2014} and in SNS junctions~\cite{larsen2020ParityProtected,schrade2022Protected, maiani2022Entangling} (and also recently in a hybrid structure~\cite{banszerus_hybrid_2024}).
When the two interferometer arms are identical and the flux is tuned to half a flux quantum, the odd-harmonic terms in the EPR cancel out, $\cos(n_\mathrm{odd} \varphi) + \cos(n_\mathrm{odd} (\varphi+\pi)) = 0$, while the even-harmonic terms remain, $\cos(n_\mathrm{even} \varphi) + \cos(n_\mathrm{even} (\varphi+\pi)) = 2\cos(n_\mathrm{even} \varphi)$. 
Because the EPR harmonics are related to Cooper pair transfers, the interferometer now only mediates transfers of even numbers of Cooper pairs between the reservoirs.
Therefore, the parity of the number of transferred Cooper pairs becomes a conserved quantity, which may be useful for designing a qubit with long coherence times.

Since both paradigmatic devices, shown in Fig.~\ref{fig:protected_qubit}, have interferometer arms that host the EPR of Eq.~\eqref{eq:theone}, the correspondence between the two implementations becomes clearer.
However, other EPRs are acceptable: the requirement is that the odd harmonics of the interferometer arms are identical and that the second harmonic exists, a flexibility which has been taken advantage of by incorporating larger SIS arrays~\cite{bell_protected_2014}.
This also shows why conventional SQUIDs with a single SIS junction in each arm cannot be used for Cooper-pair-parity protected qubits, since they do not have any higher harmonics.

\begin{figure}[htbp]
\centering
\includegraphics[width=0.8\linewidth]{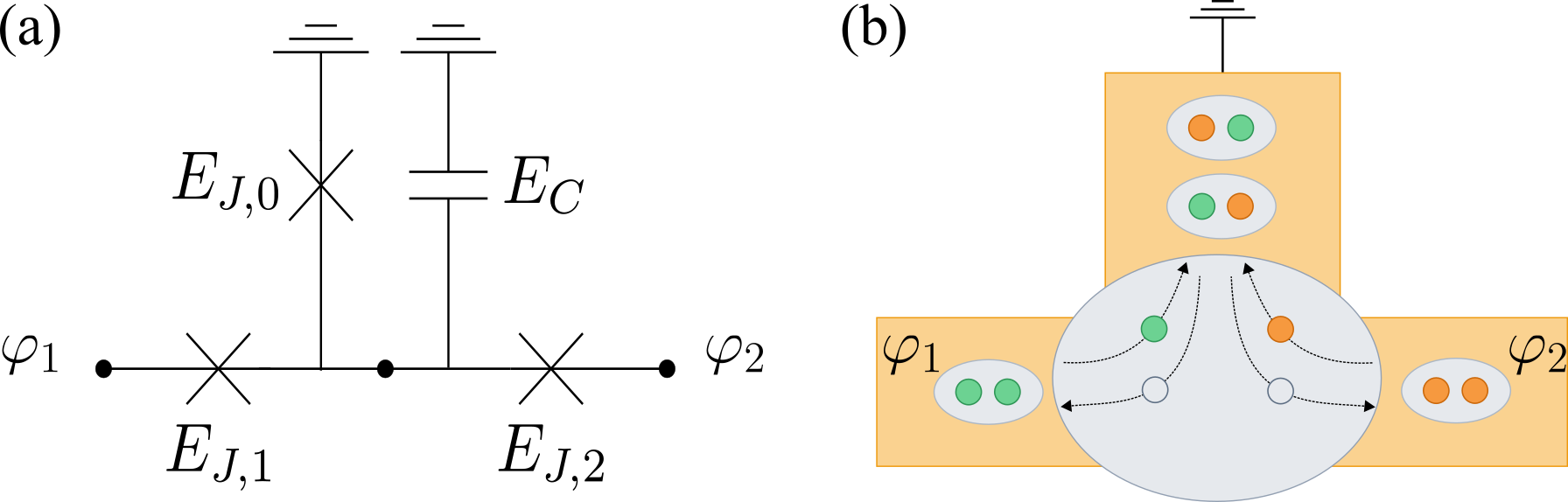}
\caption{ (a) Three terminal SIS junctions support multiplet supercurrents.
(b) An SNS junction with three terminals that allows multiplet supercurrent via non-local Andreev scattering. 
Panels adapted from Melo, et al.~\cite{melo2022Multiplet}}
\label{fig:multiplet_supercurrent}
\end{figure}

\textit{Multiplet Supercurrents:}
A multiplet supercurrent refers to the coherent transfer of more than one Cooper pair at a time between more than two contacts (Fig.~\ref{fig:multiplet_supercurrent}).
Multiplet supercurrent effects were recently observed in multiterminal SNS structures~\cite{cuevas_voltage-induced_2007,pfeffer_subgap_2014,freyn2011Production, pankratova_multiterminal_2020,zhang2023Andreev}.
Explanations based on Andreev levels that are connected to more than one terminal were proposed~\cite{freyn_production_2011,melin2023Quantum}.
Classical and quantum multiterminal SIS circuits should also exhibit multiplet supercurrent~\cite{melo_multiplet_2022}, which we briefly motivate.

The usual SIS EPR, $-E_J\cos(\varphi)$, is indicative of single Cooper pair tunneling. 
Correspondingly, terms like $\cos(n\varphi)$ indicate $n$-Cooper pair tunneling (note that Eq.~\eqref{eq:theone} has Fourier components for all $n\in \mathbb{N}$).
For a three-terminal structure (Fig.~\ref{fig:multiplet_supercurrent}), one can further have multiterminal EPR components like $\cos(n\varphi_1 + m\varphi_2)$, which indicate transfers of $n$ and $m$ Cooper pairs from terminals 1 and 2, respectively, to the grounded terminal.
Components like this indeed are present in the ground state EPR for weak and strong coupled multiterminal SIS arrays, just as for multiterminal SNS arrays.
In fact, a room temperature circuit that uses voltage-controlled oscillators to mimic the EPR of SIS junctions observed the experimental signatures of this effect, indicating the classical compatibility of this physics~\cite{arnault2022Dynamical}.

\textit{Topological Band Structures:} 
Multiterminal Andreev weak links, in which a normal region is contacted by multiple superconductors, implement a range of Hamiltonians which exhibit topological band structures~\cite{vanheck2014Single, yokoyama2015Singularities, riwar2016Multiterminal, meyer2017Nontrivial, peraltagavensky2018Berry, klees2020Microwave}.
In this context, the superconducting phase differences between the reservoirs play the role of quasimomenta. 
The topological property of the junction is characterized by topological invariants of the sub-gap states that disperse in this parameter space, and protected energy level crossings known as Weyl points appear. 
This approach provides a promising platform for engineering topological phases of matter in a controlled and tunable way, but these microscopic structures are likely difficult to fabricate experimentally.

SIS arrays provide an alternative approach to simulate such single-particle Hamiltonians~\cite{fatemi2021Weyl, peyruchat2021Transconductance, herrig2022Cooperpair}. 
In this case, instead of quantum levels derived from a microscopic degree of freedom, the SIS arrays make use of quantized collective modes.
Because SIS junction arrays are routinely fabricated and measured, we believe such circuits should more easily exhibit the topological band structures, and early experiments have so far been successful~\cite{peyruchat_spectral_2024}. 
We note in passing that this example is the only one described here that directly involves the excitations of the internal degrees of freedom between the reservoirs.

\section{Discussion}
\label{sec:discussion}
\textit{Intermediate coupling:} 
The cases considered in Sec.~\ref{sec:epr}, while instructive and useful, are the theoretical limiting cases.
At intermediate coupling, the EPRs no longer strictly follow Eq.~\eqref{eq:theone}. 
For SNS, one must account for the influence the above-gap continuum in the superconducting density of states, which itself contributes to the supercurrent and distorts the dispersion of bound state~\cite{beenakker_three_1992,beenakker1992Resonant,kurilovich_microwave_2021,fatemi_microwave_2022}. 
Additionally, violation of the Andreev approximation (e.g., if the bands disperse significantly within the energy window of $2\Delta$) results in deviations as well~\cite{kruti_impact_2024}. 
For SISIS, one must consider more carefully the dynamics of the charge on the island, such as by including additional charge states~\cite{joyez_single_1995}.
Also, we are not aware of direct mathematical correspondences when the number of series tunnel junctions is more than two~\cite{miano_hamiltonian_2023} or the SNS junction is not short~\cite{nazarov_quantum_2009}. 

\textit{Discrete levels with high supercurrent:} 
The fact that SNS junctions retain discrete Andreev levels in the strong coupling regime is special: these microscopic quantum degrees of freedom can be highly anharmonic while carrying significant supercurrent.
Such a platform for quantum information processing could combine the anharmonicity of microscopic quantum dot qubits with a degree of macroscopicity that can help with connecting to other quantum states at long range and in ultrastrong light-matter coupled devices~\cite{zazunov_andreev_2003,janvier_coherent_2015,prada_andreev_2020,pita-vidal_strong_2024,vakhtel_tunneling_2024, cheung_photon-mediated_2023}. 
Novel physics can also arise, such as the vanishing of charge dispersion in a transmon when transparency approaches unity~\cite{averin_coulomb_1999,bargerbos_observation_2020,kringhoj_suppressed_2020}. 

We highlight in particular the case of the Andreev spin qubit: the spin degree of freedom of a single quasiparticle trapped in an SNS junction, which can also determine the supercurrent~\cite{chtchelkatchev_andreev_2003,park_andreev_2017,tosi_spin-orbit_2019}.
Recently, the first experimental proof-of-concept qubit was achieved~\cite{hays_coherent_2021}, and, with these, a team has just demonstrated record-breaking two-spin coupling strengths at long distance~\cite{pita-vidal_strong_2024}. 
In contrast, a quasiparticle in the central island of an SISIS structure would reside in a continuum density of states at elevated energy $\Delta$, making quantum control of the particle or its spin difficult.

\textit{Gate tunability:}
Finally, SNS implementations with low electron density normal regions, such as semiconductors and 2d materials, afford \textit{in situ} tunability of the junction strength $E_0$ and effective transparency $\tau_\mathrm{eff}$ with an electrostatic gate voltage~\cite{jarillo-herrero_quantum_2006,heersche_bipolar_2007,larsen_semiconductor-nanowire-based_2015,heedt2021Shadowwall, vanloo2023Electrostatic}, as depicted in Fig.~\ref{fig:protected_qubit}(b).
Such \textit{in situ} tunability is a powerful tool for validating theoretical ideas, uncovering experimental realities, and parametric control, as ha been  taken advantage of in the past through magnetic flux tunability of loops containing SIS junctions~\cite{kjaergaard_superconducting_2020,blais_circuit_2021}. 
The charge tunability made possible by SNS junctions adds a complementary control axis with low cross-talk to flux-tuned circuit elements and lower power dissipation elsewhere in the cryogenic setup. 
In recent experimental works~\cite{banszerus_voltage-controlled_2024,banszerus_hybrid_2024}, it has been demonstrated that the energy-phase relation of two voltage-controlled SNS junctions in series follow Eq.~\eqref{eq:theone}, supporting a model of two conventional JJs in series~\cite{bozkurt2023DoubleFourier}.
Example demonstrations of practical applications include modulation between optimal operation points~\cite{larsen2015SemiconductorNanowireBased, casparis2018Superconducting, casparis2019Voltagecontrolled, sardashti2020VoltageTunable, strickland2023Superconducting} and activation of dynamic effects like amplification~\cite{butseraen2022gatetunable, phan2023GateTunable, sarkar2022Quantumnoiselimited,hao_kerr_2024}.
Charge control over the junction inductance also may enable novel circuit couplings for nonreciprocal devices such as the gyrator and derivative qubit concepts~\cite{leroux2022Nonreciprocal}.
In principle, gate-tunability is also possible for SISIS junctions in the weak and intermediate coupling regimes~\cite{joyez_single_1995}; however, challenges include achieving useful gate-modulation amplitudes and reliable device handling.
We also remark that protected qubit concepts, like the rhombus, often require high circuit symmetry, placing stringent requirements on Josephson junction fabrication accuracy, which is only reliable at the \% level in the state of the art~\cite{kreikebaum_improving_2020,takahashi_uniformity_2022,zheng_fabrication_2023}. 
Gate tunability can help resolve that issue for individual devices, at the cost of introducing a new noise channel.

The combination of gate-tunability and the availability of subgap spin and orbital quantum levels is appealing, and thanks to enabling materials technology from the last decade~\cite{krogstrup_epitaxy_2015,shabani_two-dimensional_2016}, SNS approaches to solid state quantum hardware have been gaining popularity.
While such devices are presently less coherent than SIS-based devices, SNS integration into superconducting quantum circuits is also much younger and, therefore, deserves more time and effort to improve. 

\section{Conclusions}

To conclude, we collected a few results that, to our knowledge, were not previously compared to describe similarities between SNS and SISIS junctions, and we showed how this comparison helps to draw connections where such junctions were separately proposed or applied towards common goals. 
In describing these relationships, we hope that we can add some clarity as both platforms are developed for sophisticated circuit design and basic physics investigations.
We anticipate that additional links will be made between these two platforms in the coming years. 

\acknowledgments 
We would like to thank A. Akhmerov, S. Diamond, N. Frattini, P. D . Kurilovich, A. L. R. Manesco, K. Nowack, B. Ramshaw, C. Schrade, L. J. Splitthoff, S. Tomarken, T. Vakhtel, B. van Heck, and J. J. Wesdorp for suggestions and discussions that improved this manuscript. 
V. F. also acknowledges P. D . Kurilovich and V. D. Kurilovich for early discussions regarding the strong coupling SISIS case. 
Research was sponsored by the Army Research Office and was accomplished under Grant W911NF2210053 Number W911NF-22-1-0053. The views and conclusions contained in this document are those of the authors and should not be interpreted as representing the official policies, either expressed or implied, of the Army Research Office or the U.S. Government. The U.S. Government is authorized to reproduce and distribute reprints for Government purposes notwithstanding any copyright notation herein.

An earlier version of this manuscript is published as part of an SPIE Proceedings~\cite{bozkurt_josephson_2023}, with digital object identifier 10.1117/12.2678477.

\bibliography{references,vf-references}

\begin{thebibliography}{101}%
\makeatletter
\providecommand \@ifxundefined [1]{%
 \@ifx{#1\undefined}
}%
\providecommand \@ifnum [1]{%
 \ifnum #1\expandafter \@firstoftwo
 \else \expandafter \@secondoftwo
 \fi
}%
\providecommand \@ifx [1]{%
 \ifx #1\expandafter \@firstoftwo
 \else \expandafter \@secondoftwo
 \fi
}%
\providecommand \natexlab [1]{#1}%
\providecommand \enquote  [1]{``#1''}%
\providecommand \bibnamefont  [1]{#1}%
\providecommand \bibfnamefont [1]{#1}%
\providecommand \citenamefont [1]{#1}%
\providecommand \href@noop [0]{\@secondoftwo}%
\providecommand \href [0]{\begingroup \@sanitize@url \@href}%
\providecommand \@href[1]{\@@startlink{#1}\@@href}%
\providecommand \@@href[1]{\endgroup#1\@@endlink}%
\providecommand \@sanitize@url [0]{\catcode `\\12\catcode `\$12\catcode
  `\&12\catcode `\#12\catcode `\^12\catcode `\_12\catcode `\%12\relax}%
\providecommand \@@startlink[1]{}%
\providecommand \@@endlink[0]{}%
\providecommand \url  [0]{\begingroup\@sanitize@url \@url }%
\providecommand \@url [1]{\endgroup\@href {#1}{\urlprefix }}%
\providecommand \urlprefix  [0]{URL }%
\providecommand \Eprint [0]{\href }%
\providecommand \doibase [0]{https://doi.org/}%
\providecommand \selectlanguage [0]{\@gobble}%
\providecommand \bibinfo  [0]{\@secondoftwo}%
\providecommand \bibfield  [0]{\@secondoftwo}%
\providecommand \translation [1]{[#1]}%
\providecommand \BibitemOpen [0]{}%
\providecommand \bibitemStop [0]{}%
\providecommand \bibitemNoStop [0]{.\EOS\space}%
\providecommand \EOS [0]{\spacefactor3000\relax}%
\providecommand \BibitemShut  [1]{\csname bibitem#1\endcsname}%
\let\auto@bib@innerbib\@empty
\bibitem [{\citenamefont {Josephson}(1962)}]{josephson_possible_1962}%
  \BibitemOpen
  \bibfield  {author} {\bibinfo {author} {\bibfnamefont {B.~D.}\ \bibnamefont
  {Josephson}},\ }\href {https://doi.org/10.1016/0031-9163(62)91369-0}
  {\bibfield  {journal} {\bibinfo  {journal} {Physics Letters}\ }\textbf
  {\bibinfo {volume} {1}},\ \bibinfo {pages} {251} (\bibinfo {year}
  {1962})}\BibitemShut {NoStop}%
\bibitem [{Note1()}]{Note1}%
  \BibitemOpen
  \bibinfo {note} {Here, we use ``normal'' to refer to any non-superconducting
  and non-insulating central region, not just metals with high electron
  density. Additionally, note that we do not consider strongly correlated
  states in the normal region like magnetism.}\BibitemShut {Stop}%
\bibitem [{\citenamefont {Tinkham}(2004)}]{tinkham_introduction_2004}%
  \BibitemOpen
  \bibfield  {author} {\bibinfo {author} {\bibfnamefont {M.}~\bibnamefont
  {Tinkham}},\ }\href@noop {} {\emph {\bibinfo {title} {Introduction to
  {Superconductivity}}}}\ (\bibinfo  {publisher} {Dover},\ \bibinfo {address}
  {New York},\ \bibinfo {year} {2004})\BibitemShut {NoStop}%
\bibitem [{\citenamefont {Hanson}\ \emph {et~al.}(2007)\citenamefont {Hanson},
  \citenamefont {Kouwenhoven}, \citenamefont {Petta}, \citenamefont {Tarucha},\
  and\ \citenamefont {Vandersypen}}]{hanson_spins_2007}%
  \BibitemOpen
  \bibfield  {author} {\bibinfo {author} {\bibfnamefont {R.}~\bibnamefont
  {Hanson}}, \bibinfo {author} {\bibfnamefont {L.~P.}\ \bibnamefont
  {Kouwenhoven}}, \bibinfo {author} {\bibfnamefont {J.~R.}\ \bibnamefont
  {Petta}}, \bibinfo {author} {\bibfnamefont {S.}~\bibnamefont {Tarucha}},\
  and\ \bibinfo {author} {\bibfnamefont {L.~M.~K.}\ \bibnamefont
  {Vandersypen}},\ }\href {https://doi.org/10.1103/RevModPhys.79.1217}
  {\bibfield  {journal} {\bibinfo  {journal} {Reviews of Modern Physics}\
  }\textbf {\bibinfo {volume} {79}},\ \bibinfo {pages} {1217} (\bibinfo {year}
  {2007})}\BibitemShut {NoStop}%
\bibitem [{\citenamefont {Prada}\ \emph {et~al.}(2020)\citenamefont {Prada},
  \citenamefont {San-Jose}, \citenamefont {de~Moor}, \citenamefont {Geresdi},
  \citenamefont {Lee}, \citenamefont {Klinovaja}, \citenamefont {Loss},
  \citenamefont {Nygård}, \citenamefont {Aguado},\ and\ \citenamefont
  {Kouwenhoven}}]{prada_andreev_2020}%
  \BibitemOpen
  \bibfield  {author} {\bibinfo {author} {\bibfnamefont {E.}~\bibnamefont
  {Prada}}, \bibinfo {author} {\bibfnamefont {P.}~\bibnamefont {San-Jose}},
  \bibinfo {author} {\bibfnamefont {M.~W.~A.}\ \bibnamefont {de~Moor}},
  \bibinfo {author} {\bibfnamefont {A.}~\bibnamefont {Geresdi}}, \bibinfo
  {author} {\bibfnamefont {E.~J.~H.}\ \bibnamefont {Lee}}, \bibinfo {author}
  {\bibfnamefont {J.}~\bibnamefont {Klinovaja}}, \bibinfo {author}
  {\bibfnamefont {D.}~\bibnamefont {Loss}}, \bibinfo {author} {\bibfnamefont
  {J.}~\bibnamefont {Nygård}}, \bibinfo {author} {\bibfnamefont
  {R.}~\bibnamefont {Aguado}},\ and\ \bibinfo {author} {\bibfnamefont {L.~P.}\
  \bibnamefont {Kouwenhoven}},\ }\href
  {https://doi.org/10.1038/s42254-020-0228-y} {\bibfield  {journal} {\bibinfo
  {journal} {Nature Reviews Physics}\ ,\ \bibinfo {pages} {1}} (\bibinfo {year}
  {2020})}\BibitemShut {NoStop}%
\bibitem [{\citenamefont {Vool}\ and\ \citenamefont
  {Devoret}(2017)}]{vool_introduction_2017}%
  \BibitemOpen
  \bibfield  {author} {\bibinfo {author} {\bibfnamefont {U.}~\bibnamefont
  {Vool}}\ and\ \bibinfo {author} {\bibfnamefont {M.}~\bibnamefont {Devoret}},\
  }\href {https://doi.org/10.1002/cta.2359} {\bibfield  {journal} {\bibinfo
  {journal} {International Journal of Circuit Theory and Applications}\
  }\textbf {\bibinfo {volume} {45}},\ \bibinfo {pages} {897} (\bibinfo {year}
  {2017})}\BibitemShut {NoStop}%
\bibitem [{\citenamefont {Beenakker}\ and\ \citenamefont {{van
  Houten}}(1992)}]{beenakker1992Resonant}%
  \BibitemOpen
  \bibfield  {author} {\bibinfo {author} {\bibfnamefont {C.~W.~J.}\
  \bibnamefont {Beenakker}}\ and\ \bibinfo {author} {\bibfnamefont
  {H.}~\bibnamefont {{van Houten}}},\ }in\ \href
  {https://doi.org/10.1007/978-3-642-77274-0_20} {\emph {\bibinfo {booktitle}
  {Single-{{Electron Tunneling}} and {{Mesoscopic Devices}}}}},\ \bibinfo
  {series and number} {Springer {{Series}} in {{Electronics}} and
  {{Photonics}}},\ \bibinfo {editor} {edited by\ \bibinfo {editor}
  {\bibfnamefont {H.}~\bibnamefont {Koch}}\ and\ \bibinfo {editor}
  {\bibfnamefont {H.}~\bibnamefont {L{\"u}bbig}}}\ (\bibinfo  {publisher}
  {{Springer}},\ \bibinfo {address} {{Berlin, Heidelberg}},\ \bibinfo {year}
  {1992})\ pp.\ \bibinfo {pages} {175--179}\BibitemShut {NoStop}%
\bibitem [{\citenamefont {Wendin}\ and\ \citenamefont
  {Shumeiko}(1996)}]{wendin1996Josephson}%
  \BibitemOpen
  \bibfield  {author} {\bibinfo {author} {\bibfnamefont {G.}~\bibnamefont
  {Wendin}}\ and\ \bibinfo {author} {\bibfnamefont {V.~S.}\ \bibnamefont
  {Shumeiko}},\ }\href {https://doi.org/10.1006/spmi.1996.0116} {\bibfield
  {journal} {\bibinfo  {journal} {Superlattices and Microstructures}\ }\textbf
  {\bibinfo {volume} {20}},\ \bibinfo {pages} {569} (\bibinfo {year}
  {1996})}\BibitemShut {NoStop}%
\bibitem [{\citenamefont {Kurilovich}\ \emph {et~al.}(2021)\citenamefont
  {Kurilovich}, \citenamefont {Kurilovich}, \citenamefont {Fatemi},
  \citenamefont {Devoret},\ and\ \citenamefont
  {Glazman}}]{kurilovich_microwave_2021}%
  \BibitemOpen
  \bibfield  {author} {\bibinfo {author} {\bibfnamefont {P.~D.}\ \bibnamefont
  {Kurilovich}}, \bibinfo {author} {\bibfnamefont {V.~D.}\ \bibnamefont
  {Kurilovich}}, \bibinfo {author} {\bibfnamefont {V.}~\bibnamefont {Fatemi}},
  \bibinfo {author} {\bibfnamefont {M.~H.}\ \bibnamefont {Devoret}},\ and\
  \bibinfo {author} {\bibfnamefont {L.~I.}\ \bibnamefont {Glazman}},\ }\href
  {https://doi.org/10.1103/PhysRevB.104.174517} {\bibfield  {journal} {\bibinfo
   {journal} {Physical Review B}\ }\textbf {\bibinfo {volume} {104}},\ \bibinfo
  {pages} {174517} (\bibinfo {year} {2021})}\BibitemShut {NoStop}%
\bibitem [{\citenamefont {Joyez}(1995)}]{joyez_single_1995}%
  \BibitemOpen
  \bibfield  {author} {\bibinfo {author} {\bibfnamefont {P.}~\bibnamefont
  {Joyez}},\ }\emph {\bibinfo {title} {The {Single} {Cooper} {Pair}
  {Transistor}: {A} {Macroscopic} {Quantum} {System}}},\ \href@noop {}
  {\bibinfo {type} {Thesis}},\ \bibinfo  {school} {Université Pierre et Marie
  Curie}, \bibinfo {address} {Paris} (\bibinfo {year} {1995})\BibitemShut
  {NoStop}%
\bibitem [{Note2()}]{Note2}%
  \BibitemOpen
  \bibinfo {note} {Even-odd fermion parity transitions are a subject of study
  for both CPT~\cite {joyez_single_1995,aumentado2004Nonequilibrium} and
  S-QD-S~\cite {martin-rodero2011Josephson,
  kadlecova_practical_2019,kurilovich_microwave_2021,bargerbos_singlet-doublet_2022,fatemi_microwave_2022},
  in part due to quasiparticle poisoning~\cite {glazman_bogoliubov_2021}.
  Unlike S-QD-S, the CPT features an inherent energy penalty $\Delta $ for odd
  parity states, making them less energetically favorable compared to even
  parity states. However, we note that, depending on specific parameter values,
  such as $\epsilon $, $\gamma _i$, and $\varphi _i$, and charging effects, the
  opposite charge parity state $(N + 1)$ can become energetically favorable in
  both systems.}\BibitemShut {Stop}%
\bibitem [{\citenamefont {Beenakker}\ and\ \citenamefont
  {Houten}(1992)}]{beenakker_resonant_1992}%
  \BibitemOpen
  \bibfield  {author} {\bibinfo {author} {\bibfnamefont {C.~W.~J.}\
  \bibnamefont {Beenakker}}\ and\ \bibinfo {author} {\bibfnamefont {H.~v.}\
  \bibnamefont {Houten}},\ }in\ \href
  {https://link.springer.com/chapter/10.1007/978-3-642-77274-0_20} {\emph
  {\bibinfo {booktitle} {Single-{Electron} {Tunneling} and {Mesoscopic}
  {Devices}}}},\ \bibinfo {series and number} {Springer {Series} in
  {Electronics} and {Photonics}}\ (\bibinfo  {publisher} {Springer, Berlin,
  Heidelberg},\ \bibinfo {year} {1992})\ pp.\ \bibinfo {pages}
  {175--179}\BibitemShut {NoStop}%
\bibitem [{\citenamefont {Friedman}\ and\ \citenamefont
  {Averin}(2002)}]{friedman2002AharonovCasherEffect}%
  \BibitemOpen
  \bibfield  {author} {\bibinfo {author} {\bibfnamefont {J.~R.}\ \bibnamefont
  {Friedman}}\ and\ \bibinfo {author} {\bibfnamefont {D.~V.}\ \bibnamefont
  {Averin}},\ }\href {https://doi.org/10.1103/PhysRevLett.88.050403} {\bibfield
   {journal} {\bibinfo  {journal} {Phys. Rev. Lett.}\ }\textbf {\bibinfo
  {volume} {88}},\ \bibinfo {pages} {050403} (\bibinfo {year}
  {2002})}\BibitemShut {NoStop}%
\bibitem [{\citenamefont {Beenakker}(1991)}]{beenakker_universal_1991}%
  \BibitemOpen
  \bibfield  {author} {\bibinfo {author} {\bibfnamefont {C.~W.~J.}\
  \bibnamefont {Beenakker}},\ }\href
  {https://doi.org/10.1103/PhysRevLett.67.3836} {\bibfield  {journal} {\bibinfo
   {journal} {Physical Review Letters}\ }\textbf {\bibinfo {volume} {67}},\
  \bibinfo {pages} {3836} (\bibinfo {year} {1991})}\BibitemShut {NoStop}%
\bibitem [{\citenamefont {Bretheau}(2013)}]{bretheau_localized_2013}%
  \BibitemOpen
  \bibfield  {author} {\bibinfo {author} {\bibfnamefont {L.}~\bibnamefont
  {Bretheau}},\ }\emph {\bibinfo {title} {Localized {Excitations} in
  {Superconducting} {Atomic} {Contacts}: {Probing} the {Andreev} {Doublet}}},\
  \href {https://pastel.archives-ouvertes.fr/pastel-00862029/document}
  {\bibinfo {type} {Thesis}},\ \bibinfo  {school} {Ecole Polytechnique}
  (\bibinfo {year} {2013})\BibitemShut {NoStop}%
\bibitem [{\citenamefont {Janvier}\ \emph {et~al.}(2015)\citenamefont
  {Janvier}, \citenamefont {Tosi}, \citenamefont {Bretheau}, \citenamefont
  {Girit}, \citenamefont {Stern}, \citenamefont {Bertet}, \citenamefont
  {Joyez}, \citenamefont {Vion}, \citenamefont {Esteve}, \citenamefont
  {Goffman}, \citenamefont {Pothier},\ and\ \citenamefont
  {Urbina}}]{janvier_coherent_2015}%
  \BibitemOpen
  \bibfield  {author} {\bibinfo {author} {\bibfnamefont {C.}~\bibnamefont
  {Janvier}}, \bibinfo {author} {\bibfnamefont {L.}~\bibnamefont {Tosi}},
  \bibinfo {author} {\bibfnamefont {L.}~\bibnamefont {Bretheau}}, \bibinfo
  {author} {\bibfnamefont {C.~O.}\ \bibnamefont {Girit}}, \bibinfo {author}
  {\bibfnamefont {M.}~\bibnamefont {Stern}}, \bibinfo {author} {\bibfnamefont
  {P.}~\bibnamefont {Bertet}}, \bibinfo {author} {\bibfnamefont
  {P.}~\bibnamefont {Joyez}}, \bibinfo {author} {\bibfnamefont
  {D.}~\bibnamefont {Vion}}, \bibinfo {author} {\bibfnamefont {D.}~\bibnamefont
  {Esteve}}, \bibinfo {author} {\bibfnamefont {M.~F.}\ \bibnamefont {Goffman}},
  \bibinfo {author} {\bibfnamefont {H.}~\bibnamefont {Pothier}},\ and\ \bibinfo
  {author} {\bibfnamefont {C.}~\bibnamefont {Urbina}},\ }\href
  {https://doi.org/10.1126/science.aab2179} {\bibfield  {journal} {\bibinfo
  {journal} {Science}\ }\textbf {\bibinfo {volume} {349}},\ \bibinfo {pages}
  {1199} (\bibinfo {year} {2015})}\BibitemShut {NoStop}%
\bibitem [{\citenamefont {Bozkurt}\ \emph {et~al.}(2023)\citenamefont
  {Bozkurt}, \citenamefont {Brookman}, \citenamefont {Fatemi},\ and\
  \citenamefont {Akhmerov}}]{bozkurt2023DoubleFourier}%
  \BibitemOpen
  \bibfield  {author} {\bibinfo {author} {\bibfnamefont {A.~M.}\ \bibnamefont
  {Bozkurt}}, \bibinfo {author} {\bibfnamefont {J.}~\bibnamefont {Brookman}},
  \bibinfo {author} {\bibfnamefont {V.}~\bibnamefont {Fatemi}},\ and\ \bibinfo
  {author} {\bibfnamefont {A.~R.}\ \bibnamefont {Akhmerov}},\ }\href
  {https://doi.org/10.21468/SciPostPhys.15.5.204} {\bibfield  {journal}
  {\bibinfo  {journal} {SciPost Physics}\ }\textbf {\bibinfo {volume} {15}},\
  \bibinfo {pages} {204} (\bibinfo {year} {2023})}\BibitemShut {NoStop}%
\bibitem [{\citenamefont {Frattini}\ \emph {et~al.}(2018)\citenamefont
  {Frattini}, \citenamefont {Sivak}, \citenamefont {Lingenfelter},
  \citenamefont {Shankar},\ and\ \citenamefont
  {Devoret}}]{frattini2018Optimizing}%
  \BibitemOpen
  \bibfield  {author} {\bibinfo {author} {\bibfnamefont {N.~E.}\ \bibnamefont
  {Frattini}}, \bibinfo {author} {\bibfnamefont {V.~V.}\ \bibnamefont {Sivak}},
  \bibinfo {author} {\bibfnamefont {A.}~\bibnamefont {Lingenfelter}}, \bibinfo
  {author} {\bibfnamefont {S.}~\bibnamefont {Shankar}},\ and\ \bibinfo {author}
  {\bibfnamefont {M.~H.}\ \bibnamefont {Devoret}},\ }\href
  {https://doi.org/10.1103/PhysRevApplied.10.054020} {\bibfield  {journal}
  {\bibinfo  {journal} {Phys. Rev. Appl.}\ }\textbf {\bibinfo {volume} {10}},\
  \bibinfo {pages} {054020} (\bibinfo {year} {2018})}\BibitemShut {NoStop}%
\bibitem [{Note3()}]{Note3}%
  \BibitemOpen
  \bibinfo {note} {An important caveat is that for exactly $\tau _\protect
  \mathrm {eff}=1$ and $\varphi =\pi $, the Josephson terms are frustrated and
  the island cannot be treated classically.}\BibitemShut {Stop}%
\bibitem [{\citenamefont {Rymarz}\ and\ \citenamefont
  {DiVincenzo}(2023)}]{rymarz2023consistent}%
  \BibitemOpen
  \bibfield  {author} {\bibinfo {author} {\bibfnamefont {M.}~\bibnamefont
  {Rymarz}}\ and\ \bibinfo {author} {\bibfnamefont {D.~P.}\ \bibnamefont
  {DiVincenzo}},\ }\href {https://doi.org/10.1103/PhysRevX.13.021017}
  {\bibfield  {journal} {\bibinfo  {journal} {Phys. Rev. X}\ }\textbf {\bibinfo
  {volume} {13}},\ \bibinfo {pages} {021017} (\bibinfo {year}
  {2023})}\BibitemShut {NoStop}%
\bibitem [{\citenamefont
  {Manucharyan}(2012)}]{manucharyan_superinductance_2012}%
  \BibitemOpen
  \bibfield  {author} {\bibinfo {author} {\bibfnamefont {V.}~\bibnamefont
  {Manucharyan}},\ }\emph {\bibinfo {title} {Superinductance}},\ \href
  {http://qulab.eng.yale.edu/documents/theses/Manucharyan,%20Vladimir%20-%20Superinductance%20(Yale,%202012).pdf}
  {Ph.D. thesis},\ \bibinfo  {school} {Yale University}, \bibinfo {address}
  {New Haven, CT} (\bibinfo {year} {2012})\BibitemShut {NoStop}%
\bibitem [{\citenamefont {Frattini}(2021)}]{frattini_three-wave_2021}%
  \BibitemOpen
  \bibfield  {author} {\bibinfo {author} {\bibfnamefont {N.}~\bibnamefont
  {Frattini}},\ }\emph {\bibinfo {title} {Three-wave {Mixing} in
  {Superconducting} {Circuits}: {Stabilizing} {Cats} with {SNAILs}}},\ \href
  {https://elischolar.library.yale.edu/gsas_dissertations/332} {Ph.D. thesis},\
  \bibinfo  {school} {Yale University} (\bibinfo {year} {2021})\BibitemShut
  {NoStop}%
\bibitem [{\citenamefont {Schrade}\ and\ \citenamefont
  {Fatemi}(2024)}]{schrade2024Dissipationless}%
  \BibitemOpen
  \bibfield  {author} {\bibinfo {author} {\bibfnamefont {C.}~\bibnamefont
  {Schrade}}\ and\ \bibinfo {author} {\bibfnamefont {V.}~\bibnamefont
  {Fatemi}},\ }\href {https://doi.org/10.1103/PhysRevApplied.21.064029}
  {\bibfield  {journal} {\bibinfo  {journal} {Phys. Rev. Appl.}\ }\textbf
  {\bibinfo {volume} {21}},\ \bibinfo {pages} {064029} (\bibinfo {year}
  {2024})}\BibitemShut {NoStop}%
\bibitem [{\citenamefont {Goldman}\ and\ \citenamefont
  {Kreisman}(1967)}]{goldman_meissner_1967}%
  \BibitemOpen
  \bibfield  {author} {\bibinfo {author} {\bibfnamefont {A.~M.}\ \bibnamefont
  {Goldman}}\ and\ \bibinfo {author} {\bibfnamefont {P.~J.}\ \bibnamefont
  {Kreisman}},\ }\href {https://doi.org/10.1103/PhysRev.164.544} {\bibfield
  {journal} {\bibinfo  {journal} {Physical Review}\ }\textbf {\bibinfo {volume}
  {164}},\ \bibinfo {pages} {544} (\bibinfo {year} {1967})}\BibitemShut
  {NoStop}%
\bibitem [{\citenamefont {Fulton}\ \emph {et~al.}(1972)\citenamefont {Fulton},
  \citenamefont {Dunkleberger},\ and\ \citenamefont
  {Dynes}}]{fulton_quantum_1972}%
  \BibitemOpen
  \bibfield  {author} {\bibinfo {author} {\bibfnamefont {T.~A.}\ \bibnamefont
  {Fulton}}, \bibinfo {author} {\bibfnamefont {L.~N.}\ \bibnamefont
  {Dunkleberger}},\ and\ \bibinfo {author} {\bibfnamefont {R.~C.}\ \bibnamefont
  {Dynes}},\ }\href {https://doi.org/10.1103/PhysRevB.6.855} {\bibfield
  {journal} {\bibinfo  {journal} {Physical Review B}\ }\textbf {\bibinfo
  {volume} {6}},\ \bibinfo {pages} {855} (\bibinfo {year} {1972})}\BibitemShut
  {NoStop}%
\bibitem [{\citenamefont {Likharev}(1986)}]{likharev_dynamics_1986}%
  \BibitemOpen
  \bibfield  {author} {\bibinfo {author} {\bibfnamefont {K.~K.}\ \bibnamefont
  {Likharev}},\ }\href@noop {} {\emph {\bibinfo {title} {Dynamics of
  {Josephson} {Junctions} and {Circuits}}}},\ \bibinfo {edition} {1st}\ ed.\
  (\bibinfo  {publisher} {CRC Press},\ \bibinfo {year} {1986})\BibitemShut
  {NoStop}%
\bibitem [{\citenamefont {Fominov}\ and\ \citenamefont
  {Mikhailov}(2022)}]{fominov2022Asymmetric}%
  \BibitemOpen
  \bibfield  {author} {\bibinfo {author} {\bibfnamefont {{\relax Ya}.~V.}\
  \bibnamefont {Fominov}}\ and\ \bibinfo {author} {\bibfnamefont {D.~S.}\
  \bibnamefont {Mikhailov}},\ }\href
  {https://doi.org/10.1103/PhysRevB.106.134514} {\bibfield  {journal} {\bibinfo
   {journal} {Phys. Rev. B}\ }\textbf {\bibinfo {volume} {106}},\ \bibinfo
  {pages} {134514} (\bibinfo {year} {2022})}\BibitemShut {NoStop}%
\bibitem [{\citenamefont {Souto}\ \emph {et~al.}(2022)\citenamefont {Souto},
  \citenamefont {Leijnse},\ and\ \citenamefont {Schrade}}]{souto2022Josephson}%
  \BibitemOpen
  \bibfield  {author} {\bibinfo {author} {\bibfnamefont {R.~S.}\ \bibnamefont
  {Souto}}, \bibinfo {author} {\bibfnamefont {M.}~\bibnamefont {Leijnse}},\
  and\ \bibinfo {author} {\bibfnamefont {C.}~\bibnamefont {Schrade}},\ }\href
  {https://doi.org/10.1103/PhysRevLett.129.267702} {\bibfield  {journal}
  {\bibinfo  {journal} {Phys. Rev. Lett.}\ }\textbf {\bibinfo {volume} {129}},\
  \bibinfo {pages} {267702} (\bibinfo {year} {2022})}\BibitemShut {NoStop}%
\bibitem [{\citenamefont {Ciaccia}\ \emph {et~al.}(2023)\citenamefont
  {Ciaccia}, \citenamefont {Haller}, \citenamefont {Drachmann}, \citenamefont
  {Lindemann}, \citenamefont {Manfra}, \citenamefont {Schrade},\ and\
  \citenamefont {Sch\"onenberger}}]{ciaccia2023Gate}%
  \BibitemOpen
  \bibfield  {author} {\bibinfo {author} {\bibfnamefont {C.}~\bibnamefont
  {Ciaccia}}, \bibinfo {author} {\bibfnamefont {R.}~\bibnamefont {Haller}},
  \bibinfo {author} {\bibfnamefont {A.~C.~C.}\ \bibnamefont {Drachmann}},
  \bibinfo {author} {\bibfnamefont {T.}~\bibnamefont {Lindemann}}, \bibinfo
  {author} {\bibfnamefont {M.~J.}\ \bibnamefont {Manfra}}, \bibinfo {author}
  {\bibfnamefont {C.}~\bibnamefont {Schrade}},\ and\ \bibinfo {author}
  {\bibfnamefont {C.}~\bibnamefont {Sch\"onenberger}},\ }\href
  {https://doi.org/10.1103/PhysRevResearch.5.033131} {\bibfield  {journal}
  {\bibinfo  {journal} {Phys. Rev. Res.}\ }\textbf {\bibinfo {volume} {5}},\
  \bibinfo {pages} {033131} (\bibinfo {year} {2023})}\BibitemShut {NoStop}%
\bibitem [{\citenamefont {Valentini}\ \emph {et~al.}(2024)\citenamefont
  {Valentini}, \citenamefont {Sagi}, \citenamefont {Baghumyan}, \citenamefont
  {de~Gijsel}, \citenamefont {Jung}, \citenamefont {Calcaterra}, \citenamefont
  {Ballabio}, \citenamefont {Aguilera~Servin}, \citenamefont {Aggarwal},
  \citenamefont {Janik}, \citenamefont {Adletzberger}, \citenamefont
  {Seoane~Souto}, \citenamefont {Leijnse}, \citenamefont {Danon}, \citenamefont
  {Schrade}, \citenamefont {Bakkers}, \citenamefont {Chrastina}, \citenamefont
  {Isella},\ and\ \citenamefont {Katsaros}}]{valentini2023Radio}%
  \BibitemOpen
  \bibfield  {author} {\bibinfo {author} {\bibfnamefont {M.}~\bibnamefont
  {Valentini}}, \bibinfo {author} {\bibfnamefont {O.}~\bibnamefont {Sagi}},
  \bibinfo {author} {\bibfnamefont {L.}~\bibnamefont {Baghumyan}}, \bibinfo
  {author} {\bibfnamefont {T.}~\bibnamefont {de~Gijsel}}, \bibinfo {author}
  {\bibfnamefont {J.}~\bibnamefont {Jung}}, \bibinfo {author} {\bibfnamefont
  {S.}~\bibnamefont {Calcaterra}}, \bibinfo {author} {\bibfnamefont
  {A.}~\bibnamefont {Ballabio}}, \bibinfo {author} {\bibfnamefont
  {J.}~\bibnamefont {Aguilera~Servin}}, \bibinfo {author} {\bibfnamefont
  {K.}~\bibnamefont {Aggarwal}}, \bibinfo {author} {\bibfnamefont
  {M.}~\bibnamefont {Janik}}, \bibinfo {author} {\bibfnamefont
  {T.}~\bibnamefont {Adletzberger}}, \bibinfo {author} {\bibfnamefont
  {R.}~\bibnamefont {Seoane~Souto}}, \bibinfo {author} {\bibfnamefont
  {M.}~\bibnamefont {Leijnse}}, \bibinfo {author} {\bibfnamefont
  {J.}~\bibnamefont {Danon}}, \bibinfo {author} {\bibfnamefont
  {C.}~\bibnamefont {Schrade}}, \bibinfo {author} {\bibfnamefont
  {E.}~\bibnamefont {Bakkers}}, \bibinfo {author} {\bibfnamefont
  {D.}~\bibnamefont {Chrastina}}, \bibinfo {author} {\bibfnamefont
  {G.}~\bibnamefont {Isella}},\ and\ \bibinfo {author} {\bibfnamefont
  {G.}~\bibnamefont {Katsaros}},\ }\href
  {https://doi.org/10.1038/s41467-023-44114-0} {\bibfield  {journal} {\bibinfo
  {journal} {Nat Commun}\ }\textbf {\bibinfo {volume} {15}},\ \bibinfo {pages}
  {169} (\bibinfo {year} {2024})}\BibitemShut {NoStop}%
\bibitem [{\citenamefont {Gladchenko}\ \emph {et~al.}(2009)\citenamefont
  {Gladchenko}, \citenamefont {Olaya}, \citenamefont {Dupont-Ferrier},
  \citenamefont {Douçot}, \citenamefont {Ioffe},\ and\ \citenamefont
  {Gershenson}}]{gladchenko_superconducting_2009}%
  \BibitemOpen
  \bibfield  {author} {\bibinfo {author} {\bibfnamefont {S.}~\bibnamefont
  {Gladchenko}}, \bibinfo {author} {\bibfnamefont {D.}~\bibnamefont {Olaya}},
  \bibinfo {author} {\bibfnamefont {E.}~\bibnamefont {Dupont-Ferrier}},
  \bibinfo {author} {\bibfnamefont {B.}~\bibnamefont {Douçot}}, \bibinfo
  {author} {\bibfnamefont {L.~B.}\ \bibnamefont {Ioffe}},\ and\ \bibinfo
  {author} {\bibfnamefont {M.~E.}\ \bibnamefont {Gershenson}},\ }\href
  {https://doi.org/10.1038/nphys1151} {\bibfield  {journal} {\bibinfo
  {journal} {Nature Physics}\ }\textbf {\bibinfo {volume} {5}},\ \bibinfo
  {pages} {48} (\bibinfo {year} {2009})}\BibitemShut {NoStop}%
\bibitem [{\citenamefont {Bell}\ \emph {et~al.}(2014)\citenamefont {Bell},
  \citenamefont {Paramanandam}, \citenamefont {Ioffe},\ and\ \citenamefont
  {Gershenson}}]{bell_protected_2014}%
  \BibitemOpen
  \bibfield  {author} {\bibinfo {author} {\bibfnamefont {M.~T.}\ \bibnamefont
  {Bell}}, \bibinfo {author} {\bibfnamefont {J.}~\bibnamefont {Paramanandam}},
  \bibinfo {author} {\bibfnamefont {L.~B.}\ \bibnamefont {Ioffe}},\ and\
  \bibinfo {author} {\bibfnamefont {M.~E.}\ \bibnamefont {Gershenson}},\ }\href
  {https://doi.org/10.1103/PhysRevLett.112.167001} {\bibfield  {journal}
  {\bibinfo  {journal} {Physical Review Letters}\ }\textbf {\bibinfo {volume}
  {112}},\ \bibinfo {pages} {167001} (\bibinfo {year} {2014})}\BibitemShut
  {NoStop}%
\bibitem [{\citenamefont {Larsen}\ \emph {et~al.}(2020)\citenamefont {Larsen},
  \citenamefont {Gershenson}, \citenamefont {Casparis}, \citenamefont
  {Kringh{\o}j}, \citenamefont {Pearson}, \citenamefont {McNeil}, \citenamefont
  {Kuemmeth}, \citenamefont {Krogstrup}, \citenamefont {Petersson},\ and\
  \citenamefont {Marcus}}]{larsen2020ParityProtected}%
  \BibitemOpen
  \bibfield  {author} {\bibinfo {author} {\bibfnamefont {T.~W.}\ \bibnamefont
  {Larsen}}, \bibinfo {author} {\bibfnamefont {M.~E.}\ \bibnamefont
  {Gershenson}}, \bibinfo {author} {\bibfnamefont {L.}~\bibnamefont
  {Casparis}}, \bibinfo {author} {\bibfnamefont {A.}~\bibnamefont
  {Kringh{\o}j}}, \bibinfo {author} {\bibfnamefont {N.~J.}\ \bibnamefont
  {Pearson}}, \bibinfo {author} {\bibfnamefont {R.~P.~G.}\ \bibnamefont
  {McNeil}}, \bibinfo {author} {\bibfnamefont {F.}~\bibnamefont {Kuemmeth}},
  \bibinfo {author} {\bibfnamefont {P.}~\bibnamefont {Krogstrup}}, \bibinfo
  {author} {\bibfnamefont {K.~D.}\ \bibnamefont {Petersson}},\ and\ \bibinfo
  {author} {\bibfnamefont {C.~M.}\ \bibnamefont {Marcus}},\ }\href
  {https://doi.org/10.1103/PhysRevLett.125.056801} {\bibfield  {journal}
  {\bibinfo  {journal} {Phys. Rev. Lett.}\ }\textbf {\bibinfo {volume} {125}},\
  \bibinfo {pages} {056801} (\bibinfo {year} {2020})}\BibitemShut {NoStop}%
\bibitem [{\citenamefont {Schrade}\ \emph {et~al.}(2022)\citenamefont
  {Schrade}, \citenamefont {Marcus},\ and\ \citenamefont
  {Gyenis}}]{schrade2022Protected}%
  \BibitemOpen
  \bibfield  {author} {\bibinfo {author} {\bibfnamefont {C.}~\bibnamefont
  {Schrade}}, \bibinfo {author} {\bibfnamefont {C.~M.}\ \bibnamefont
  {Marcus}},\ and\ \bibinfo {author} {\bibfnamefont {A.}~\bibnamefont
  {Gyenis}},\ }\href {https://doi.org/10.1103/PRXQuantum.3.030303} {\bibfield
  {journal} {\bibinfo  {journal} {PRX Quantum}\ }\textbf {\bibinfo {volume}
  {3}},\ \bibinfo {pages} {030303} (\bibinfo {year} {2022})}\BibitemShut
  {NoStop}%
\bibitem [{\citenamefont {Maiani}\ \emph {et~al.}(2022)\citenamefont {Maiani},
  \citenamefont {Kjaergaard},\ and\ \citenamefont
  {Schrade}}]{maiani2022Entangling}%
  \BibitemOpen
  \bibfield  {author} {\bibinfo {author} {\bibfnamefont {A.}~\bibnamefont
  {Maiani}}, \bibinfo {author} {\bibfnamefont {M.}~\bibnamefont {Kjaergaard}},\
  and\ \bibinfo {author} {\bibfnamefont {C.}~\bibnamefont {Schrade}},\ }\href
  {https://doi.org/10.1103/PRXQuantum.3.030329} {\bibfield  {journal} {\bibinfo
   {journal} {PRX Quantum}\ }\textbf {\bibinfo {volume} {3}},\ \bibinfo {pages}
  {030329} (\bibinfo {year} {2022})}\BibitemShut {NoStop}%
\bibitem [{\citenamefont {Banszerus}\ \emph
  {et~al.}(2024{\natexlab{a}})\citenamefont {Banszerus}, \citenamefont
  {Andersson}, \citenamefont {Marshall}, \citenamefont {Lindemann},
  \citenamefont {Manfra}, \citenamefont {Marcus},\ and\ \citenamefont
  {Vaitiekėnas}}]{banszerus_hybrid_2024}%
  \BibitemOpen
  \bibfield  {author} {\bibinfo {author} {\bibfnamefont {L.}~\bibnamefont
  {Banszerus}}, \bibinfo {author} {\bibfnamefont {C.~W.}\ \bibnamefont
  {Andersson}}, \bibinfo {author} {\bibfnamefont {W.}~\bibnamefont {Marshall}},
  \bibinfo {author} {\bibfnamefont {T.}~\bibnamefont {Lindemann}}, \bibinfo
  {author} {\bibfnamefont {M.~J.}\ \bibnamefont {Manfra}}, \bibinfo {author}
  {\bibfnamefont {C.~M.}\ \bibnamefont {Marcus}},\ and\ \bibinfo {author}
  {\bibfnamefont {S.}~\bibnamefont {Vaitiekėnas}},\ }\href
  {https://doi.org/10.48550/arXiv.2406.20082} {\bibinfo {title} {The hybrid
  {Josephson} rhombus: {A} superconducting element with tailored current-phase
  relation}} (\bibinfo {year} {2024}{\natexlab{a}}),\ \bibinfo {note}
  {arXiv:2406.20082 [cond-mat]}\BibitemShut {NoStop}%
\bibitem [{\citenamefont {Melo}\ \emph
  {et~al.}(2022{\natexlab{a}})\citenamefont {Melo}, \citenamefont {Fatemi},\
  and\ \citenamefont {Akhmerov}}]{melo2022Multiplet}%
  \BibitemOpen
  \bibfield  {author} {\bibinfo {author} {\bibfnamefont {A.}~\bibnamefont
  {Melo}}, \bibinfo {author} {\bibfnamefont {V.}~\bibnamefont {Fatemi}},\ and\
  \bibinfo {author} {\bibfnamefont {A.~R.}\ \bibnamefont {Akhmerov}},\ }\href
  {https://doi.org/10.21468/SciPostPhys.12.1.017} {\bibfield  {journal}
  {\bibinfo  {journal} {SciPost Physics}\ }\textbf {\bibinfo {volume} {12}},\
  \bibinfo {pages} {017} (\bibinfo {year} {2022}{\natexlab{a}})}\BibitemShut
  {NoStop}%
\bibitem [{\citenamefont {Cuevas}\ and\ \citenamefont
  {Pothier}(2007)}]{cuevas_voltage-induced_2007}%
  \BibitemOpen
  \bibfield  {author} {\bibinfo {author} {\bibfnamefont {J.~C.}\ \bibnamefont
  {Cuevas}}\ and\ \bibinfo {author} {\bibfnamefont {H.}~\bibnamefont
  {Pothier}},\ }\href {https://doi.org/10.1103/PhysRevB.75.174513} {\bibfield
  {journal} {\bibinfo  {journal} {Physical Review B}\ }\textbf {\bibinfo
  {volume} {75}},\ \bibinfo {pages} {174513} (\bibinfo {year}
  {2007})}\BibitemShut {NoStop}%
\bibitem [{\citenamefont {Pfeffer}\ \emph {et~al.}(2014)\citenamefont
  {Pfeffer}, \citenamefont {Duvauchelle}, \citenamefont {Courtois},
  \citenamefont {M\'elin}, \citenamefont {Feinberg},\ and\ \citenamefont
  {Lefloch}}]{pfeffer_subgap_2014}%
  \BibitemOpen
  \bibfield  {author} {\bibinfo {author} {\bibfnamefont {A.~H.}\ \bibnamefont
  {Pfeffer}}, \bibinfo {author} {\bibfnamefont {J.~E.}\ \bibnamefont
  {Duvauchelle}}, \bibinfo {author} {\bibfnamefont {H.}~\bibnamefont
  {Courtois}}, \bibinfo {author} {\bibfnamefont {R.}~\bibnamefont {M\'elin}},
  \bibinfo {author} {\bibfnamefont {D.}~\bibnamefont {Feinberg}},\ and\
  \bibinfo {author} {\bibfnamefont {F.}~\bibnamefont {Lefloch}},\ }\href
  {https://doi.org/10.1103/PhysRevB.90.075401} {\bibfield  {journal} {\bibinfo
  {journal} {Phys. Rev. B}\ }\textbf {\bibinfo {volume} {90}},\ \bibinfo
  {pages} {075401} (\bibinfo {year} {2014})}\BibitemShut {NoStop}%
\bibitem [{\citenamefont {Freyn}\ \emph
  {et~al.}(2011{\natexlab{a}})\citenamefont {Freyn}, \citenamefont {Dou{\c
  c}ot}, \citenamefont {Feinberg},\ and\ \citenamefont
  {M{\'e}lin}}]{freyn2011Production}%
  \BibitemOpen
  \bibfield  {author} {\bibinfo {author} {\bibfnamefont {A.}~\bibnamefont
  {Freyn}}, \bibinfo {author} {\bibfnamefont {B.}~\bibnamefont {Dou{\c c}ot}},
  \bibinfo {author} {\bibfnamefont {D.}~\bibnamefont {Feinberg}},\ and\
  \bibinfo {author} {\bibfnamefont {R.}~\bibnamefont {M{\'e}lin}},\ }\href
  {https://doi.org/10.1103/PhysRevLett.106.257005} {\bibfield  {journal}
  {\bibinfo  {journal} {Phys. Rev. Lett.}\ }\textbf {\bibinfo {volume} {106}},\
  \bibinfo {pages} {257005} (\bibinfo {year} {2011}{\natexlab{a}})}\BibitemShut
  {NoStop}%
\bibitem [{\citenamefont {Pankratova}\ \emph {et~al.}(2020)\citenamefont
  {Pankratova}, \citenamefont {Lee}, \citenamefont {Kuzmin}, \citenamefont
  {Wickramasinghe}, \citenamefont {Mayer}, \citenamefont {Yuan}, \citenamefont
  {Vavilov}, \citenamefont {Shabani},\ and\ \citenamefont
  {Manucharyan}}]{pankratova_multiterminal_2020}%
  \BibitemOpen
  \bibfield  {author} {\bibinfo {author} {\bibfnamefont {N.}~\bibnamefont
  {Pankratova}}, \bibinfo {author} {\bibfnamefont {H.}~\bibnamefont {Lee}},
  \bibinfo {author} {\bibfnamefont {R.}~\bibnamefont {Kuzmin}}, \bibinfo
  {author} {\bibfnamefont {K.}~\bibnamefont {Wickramasinghe}}, \bibinfo
  {author} {\bibfnamefont {W.}~\bibnamefont {Mayer}}, \bibinfo {author}
  {\bibfnamefont {J.}~\bibnamefont {Yuan}}, \bibinfo {author} {\bibfnamefont
  {M.~G.}\ \bibnamefont {Vavilov}}, \bibinfo {author} {\bibfnamefont
  {J.}~\bibnamefont {Shabani}},\ and\ \bibinfo {author} {\bibfnamefont {V.~E.}\
  \bibnamefont {Manucharyan}},\ }\href
  {https://doi.org/10.1103/PhysRevX.10.031051} {\bibfield  {journal} {\bibinfo
  {journal} {Physical Review X}\ }\textbf {\bibinfo {volume} {10}},\ \bibinfo
  {pages} {031051} (\bibinfo {year} {2020})}\BibitemShut {NoStop}%
\bibitem [{\citenamefont {Zhang}\ \emph {et~al.}(2023)\citenamefont {Zhang},
  \citenamefont {Rashid}, \citenamefont {Ahari}, \citenamefont {Zhang},
  \citenamefont {Ananthanarayanan}, \citenamefont {Xiao}, \citenamefont {{de
  Coster}}, \citenamefont {Gilbert}, \citenamefont {Samarth},\ and\
  \citenamefont {Kayyalha}}]{zhang2023Andreev}%
  \BibitemOpen
  \bibfield  {author} {\bibinfo {author} {\bibfnamefont {F.}~\bibnamefont
  {Zhang}}, \bibinfo {author} {\bibfnamefont {A.~S.}\ \bibnamefont {Rashid}},
  \bibinfo {author} {\bibfnamefont {M.~T.}\ \bibnamefont {Ahari}}, \bibinfo
  {author} {\bibfnamefont {W.}~\bibnamefont {Zhang}}, \bibinfo {author}
  {\bibfnamefont {K.~M.}\ \bibnamefont {Ananthanarayanan}}, \bibinfo {author}
  {\bibfnamefont {R.}~\bibnamefont {Xiao}}, \bibinfo {author} {\bibfnamefont
  {G.~J.}\ \bibnamefont {{de Coster}}}, \bibinfo {author} {\bibfnamefont
  {M.~J.}\ \bibnamefont {Gilbert}}, \bibinfo {author} {\bibfnamefont
  {N.}~\bibnamefont {Samarth}},\ and\ \bibinfo {author} {\bibfnamefont
  {M.}~\bibnamefont {Kayyalha}},\ }\href
  {https://doi.org/10.1103/PhysRevB.107.L140503} {\bibfield  {journal}
  {\bibinfo  {journal} {Phys. Rev. B}\ }\textbf {\bibinfo {volume} {107}},\
  \bibinfo {pages} {L140503} (\bibinfo {year} {2023})}\BibitemShut {NoStop}%
\bibitem [{\citenamefont {Freyn}\ \emph
  {et~al.}(2011{\natexlab{b}})\citenamefont {Freyn}, \citenamefont {Douçot},
  \citenamefont {Feinberg},\ and\ \citenamefont
  {Mélin}}]{freyn_production_2011}%
  \BibitemOpen
  \bibfield  {author} {\bibinfo {author} {\bibfnamefont {A.}~\bibnamefont
  {Freyn}}, \bibinfo {author} {\bibfnamefont {B.}~\bibnamefont {Douçot}},
  \bibinfo {author} {\bibfnamefont {D.}~\bibnamefont {Feinberg}},\ and\
  \bibinfo {author} {\bibfnamefont {R.}~\bibnamefont {Mélin}},\ }\href
  {https://doi.org/10.1103/PhysRevLett.106.257005} {\bibfield  {journal}
  {\bibinfo  {journal} {Physical Review Letters}\ }\textbf {\bibinfo {volume}
  {106}},\ \bibinfo {pages} {257005} (\bibinfo {year}
  {2011}{\natexlab{b}})}\BibitemShut {NoStop}%
\bibitem [{\citenamefont {M{\'e}lin}\ and\ \citenamefont
  {Feinberg}(2023)}]{melin2023Quantum}%
  \BibitemOpen
  \bibfield  {author} {\bibinfo {author} {\bibfnamefont {R.}~\bibnamefont
  {M{\'e}lin}}\ and\ \bibinfo {author} {\bibfnamefont {D.}~\bibnamefont
  {Feinberg}},\ }\href {https://doi.org/10.1103/PhysRevB.107.L161405}
  {\bibfield  {journal} {\bibinfo  {journal} {Phys. Rev. B}\ }\textbf {\bibinfo
  {volume} {107}},\ \bibinfo {pages} {L161405} (\bibinfo {year}
  {2023})}\BibitemShut {NoStop}%
\bibitem [{\citenamefont {Melo}\ \emph
  {et~al.}(2022{\natexlab{b}})\citenamefont {Melo}, \citenamefont {Fatemi},\
  and\ \citenamefont {Akhmerov}}]{melo_multiplet_2022}%
  \BibitemOpen
  \bibfield  {author} {\bibinfo {author} {\bibfnamefont {A.}~\bibnamefont
  {Melo}}, \bibinfo {author} {\bibfnamefont {V.}~\bibnamefont {Fatemi}},\ and\
  \bibinfo {author} {\bibfnamefont {A.}~\bibnamefont {Akhmerov}},\ }\href
  {https://doi.org/10.21468/SciPostPhys.12.1.017} {\bibfield  {journal}
  {\bibinfo  {journal} {SciPost Physics}\ }\textbf {\bibinfo {volume} {12}},\
  \bibinfo {pages} {017} (\bibinfo {year} {2022}{\natexlab{b}})}\BibitemShut
  {NoStop}%
\bibitem [{\citenamefont {Arnault}\ \emph {et~al.}(2022)\citenamefont
  {Arnault}, \citenamefont {Idris}, \citenamefont {McConnell}, \citenamefont
  {Zhao}, \citenamefont {Larson}, \citenamefont {Watanabe}, \citenamefont
  {Taniguchi}, \citenamefont {Finkelstein},\ and\ \citenamefont
  {Amet}}]{arnault2022Dynamical}%
  \BibitemOpen
  \bibfield  {author} {\bibinfo {author} {\bibfnamefont {E.~G.}\ \bibnamefont
  {Arnault}}, \bibinfo {author} {\bibfnamefont {S.}~\bibnamefont {Idris}},
  \bibinfo {author} {\bibfnamefont {A.}~\bibnamefont {McConnell}}, \bibinfo
  {author} {\bibfnamefont {L.}~\bibnamefont {Zhao}}, \bibinfo {author}
  {\bibfnamefont {T.~F.}\ \bibnamefont {Larson}}, \bibinfo {author}
  {\bibfnamefont {K.}~\bibnamefont {Watanabe}}, \bibinfo {author}
  {\bibfnamefont {T.}~\bibnamefont {Taniguchi}}, \bibinfo {author}
  {\bibfnamefont {G.}~\bibnamefont {Finkelstein}},\ and\ \bibinfo {author}
  {\bibfnamefont {F.}~\bibnamefont {Amet}},\ }\href
  {https://doi.org/10.1021/acs.nanolett.2c01999} {\bibfield  {journal}
  {\bibinfo  {journal} {Nano Lett.}\ }\textbf {\bibinfo {volume} {22}},\
  \bibinfo {pages} {7073} (\bibinfo {year} {2022})}\BibitemShut {NoStop}%
\bibitem [{\citenamefont {{van Heck}}\ \emph {et~al.}(2014)\citenamefont {{van
  Heck}}, \citenamefont {Mi},\ and\ \citenamefont
  {Akhmerov}}]{vanheck2014Single}%
  \BibitemOpen
  \bibfield  {author} {\bibinfo {author} {\bibfnamefont {B.}~\bibnamefont {{van
  Heck}}}, \bibinfo {author} {\bibfnamefont {S.}~\bibnamefont {Mi}},\ and\
  \bibinfo {author} {\bibfnamefont {A.~R.}\ \bibnamefont {Akhmerov}},\ }\href
  {https://doi.org/10.1103/PhysRevB.90.155450} {\bibfield  {journal} {\bibinfo
  {journal} {Phys. Rev. B}\ }\textbf {\bibinfo {volume} {90}},\ \bibinfo
  {pages} {155450} (\bibinfo {year} {2014})}\BibitemShut {NoStop}%
\bibitem [{\citenamefont {Yokoyama}\ and\ \citenamefont
  {Nazarov}(2015)}]{yokoyama2015Singularities}%
  \BibitemOpen
  \bibfield  {author} {\bibinfo {author} {\bibfnamefont {T.}~\bibnamefont
  {Yokoyama}}\ and\ \bibinfo {author} {\bibfnamefont {Y.~V.}\ \bibnamefont
  {Nazarov}},\ }\href {https://doi.org/10.1103/PhysRevB.92.155437} {\bibfield
  {journal} {\bibinfo  {journal} {Phys. Rev. B}\ }\textbf {\bibinfo {volume}
  {92}},\ \bibinfo {pages} {155437} (\bibinfo {year} {2015})}\BibitemShut
  {NoStop}%
\bibitem [{\citenamefont {Riwar}\ \emph {et~al.}(2016)\citenamefont {Riwar},
  \citenamefont {Houzet}, \citenamefont {Meyer},\ and\ \citenamefont
  {Nazarov}}]{riwar2016Multiterminal}%
  \BibitemOpen
  \bibfield  {author} {\bibinfo {author} {\bibfnamefont {R.-P.}\ \bibnamefont
  {Riwar}}, \bibinfo {author} {\bibfnamefont {M.}~\bibnamefont {Houzet}},
  \bibinfo {author} {\bibfnamefont {J.~S.}\ \bibnamefont {Meyer}},\ and\
  \bibinfo {author} {\bibfnamefont {Y.~V.}\ \bibnamefont {Nazarov}},\ }\href
  {https://doi.org/10.1038/ncomms11167} {\bibfield  {journal} {\bibinfo
  {journal} {Nat. Commun}\ }\textbf {\bibinfo {volume} {7}},\ \bibinfo {pages}
  {11167} (\bibinfo {year} {2016})}\BibitemShut {NoStop}%
\bibitem [{\citenamefont {Meyer}\ and\ \citenamefont
  {Houzet}(2017)}]{meyer2017Nontrivial}%
  \BibitemOpen
  \bibfield  {author} {\bibinfo {author} {\bibfnamefont {J.~S.}\ \bibnamefont
  {Meyer}}\ and\ \bibinfo {author} {\bibfnamefont {M.}~\bibnamefont {Houzet}},\
  }\href {https://doi.org/10.1103/PhysRevLett.119.136807} {\bibfield  {journal}
  {\bibinfo  {journal} {Phys. Rev. Lett.}\ }\textbf {\bibinfo {volume} {119}},\
  \bibinfo {pages} {136807} (\bibinfo {year} {2017})}\BibitemShut {NoStop}%
\bibitem [{\citenamefont {Peralta~Gavensky}\ \emph {et~al.}(2018)\citenamefont
  {Peralta~Gavensky}, \citenamefont {Usaj}, \citenamefont {Feinberg},\ and\
  \citenamefont {Balseiro}}]{peraltagavensky2018Berry}%
  \BibitemOpen
  \bibfield  {author} {\bibinfo {author} {\bibfnamefont {L.}~\bibnamefont
  {Peralta~Gavensky}}, \bibinfo {author} {\bibfnamefont {G.}~\bibnamefont
  {Usaj}}, \bibinfo {author} {\bibfnamefont {D.}~\bibnamefont {Feinberg}},\
  and\ \bibinfo {author} {\bibfnamefont {C.~A.}\ \bibnamefont {Balseiro}},\
  }\href {https://doi.org/10.1103/PhysRevB.97.220505} {\bibfield  {journal}
  {\bibinfo  {journal} {Phys. Rev. B}\ }\textbf {\bibinfo {volume} {97}},\
  \bibinfo {pages} {220505} (\bibinfo {year} {2018})}\BibitemShut {NoStop}%
\bibitem [{\citenamefont {Klees}\ \emph {et~al.}(2020)\citenamefont {Klees},
  \citenamefont {Rastelli}, \citenamefont {Cuevas},\ and\ \citenamefont
  {Belzig}}]{klees2020Microwave}%
  \BibitemOpen
  \bibfield  {author} {\bibinfo {author} {\bibfnamefont {R.~L.}\ \bibnamefont
  {Klees}}, \bibinfo {author} {\bibfnamefont {G.}~\bibnamefont {Rastelli}},
  \bibinfo {author} {\bibfnamefont {J.~C.}\ \bibnamefont {Cuevas}},\ and\
  \bibinfo {author} {\bibfnamefont {W.}~\bibnamefont {Belzig}},\ }\href
  {https://doi.org/10.1103/PhysRevLett.124.197002} {\bibfield  {journal}
  {\bibinfo  {journal} {Phys. Rev. Lett.}\ }\textbf {\bibinfo {volume} {124}},\
  \bibinfo {pages} {197002} (\bibinfo {year} {2020})}\BibitemShut {NoStop}%
\bibitem [{\citenamefont {Fatemi}\ \emph {et~al.}(2021)\citenamefont {Fatemi},
  \citenamefont {Akhmerov},\ and\ \citenamefont {Bretheau}}]{fatemi2021Weyl}%
  \BibitemOpen
  \bibfield  {author} {\bibinfo {author} {\bibfnamefont {V.}~\bibnamefont
  {Fatemi}}, \bibinfo {author} {\bibfnamefont {A.~R.}\ \bibnamefont
  {Akhmerov}},\ and\ \bibinfo {author} {\bibfnamefont {L.}~\bibnamefont
  {Bretheau}},\ }\href {https://doi.org/10.1103/PhysRevResearch.3.013288}
  {\bibfield  {journal} {\bibinfo  {journal} {Phys. Rev. Res.}\ }\textbf
  {\bibinfo {volume} {3}},\ \bibinfo {pages} {013288} (\bibinfo {year}
  {2021})}\BibitemShut {NoStop}%
\bibitem [{\citenamefont {Peyruchat}\ \emph {et~al.}(2021)\citenamefont
  {Peyruchat}, \citenamefont {Griesmar}, \citenamefont {Pillet},\ and\
  \citenamefont {Girit}}]{peyruchat2021Transconductance}%
  \BibitemOpen
  \bibfield  {author} {\bibinfo {author} {\bibfnamefont {L.}~\bibnamefont
  {Peyruchat}}, \bibinfo {author} {\bibfnamefont {J.}~\bibnamefont {Griesmar}},
  \bibinfo {author} {\bibfnamefont {J.-D.}\ \bibnamefont {Pillet}},\ and\
  \bibinfo {author} {\bibfnamefont {{\c C}.~{\"O}.}\ \bibnamefont {Girit}},\
  }\href {https://doi.org/10.1103/PhysRevResearch.3.013289} {\bibfield
  {journal} {\bibinfo  {journal} {Phys. Rev. Res.}\ }\textbf {\bibinfo {volume}
  {3}},\ \bibinfo {pages} {013289} (\bibinfo {year} {2021})}\BibitemShut
  {NoStop}%
\bibitem [{\citenamefont {Herrig}\ and\ \citenamefont
  {Riwar}(2022)}]{herrig2022Cooperpair}%
  \BibitemOpen
  \bibfield  {author} {\bibinfo {author} {\bibfnamefont {T.}~\bibnamefont
  {Herrig}}\ and\ \bibinfo {author} {\bibfnamefont {R.-P.}\ \bibnamefont
  {Riwar}},\ }\href {https://doi.org/10.1103/PhysRevResearch.4.013038}
  {\bibfield  {journal} {\bibinfo  {journal} {Phys. Rev. Res.}\ }\textbf
  {\bibinfo {volume} {4}},\ \bibinfo {pages} {013038} (\bibinfo {year}
  {2022})}\BibitemShut {NoStop}%
\bibitem [{\citenamefont {Peyruchat}\ \emph {et~al.}(2024)\citenamefont
  {Peyruchat}, \citenamefont {Rodriguez}, \citenamefont {Smirr}, \citenamefont
  {Leone},\ and\ \citenamefont {Girit}}]{peyruchat_spectral_2024}%
  \BibitemOpen
  \bibfield  {author} {\bibinfo {author} {\bibfnamefont {L.}~\bibnamefont
  {Peyruchat}}, \bibinfo {author} {\bibfnamefont {R.~H.}\ \bibnamefont
  {Rodriguez}}, \bibinfo {author} {\bibfnamefont {J.-L.}\ \bibnamefont
  {Smirr}}, \bibinfo {author} {\bibfnamefont {R.}~\bibnamefont {Leone}},\ and\
  \bibinfo {author} {\bibfnamefont {C.~O.}\ \bibnamefont {Girit}},\ }\href
  {https://doi.org/10.48550/arXiv.2401.10876} {\bibinfo {title} {Spectral
  signatures of non-trivial topology in a superconducting circuit}} (\bibinfo
  {year} {2024})\BibitemShut {NoStop}%
\bibitem [{\citenamefont {Beenakker}(1992)}]{beenakker_three_1992}%
  \BibitemOpen
  \bibfield  {author} {\bibinfo {author} {\bibfnamefont {C.~W.~J.}\
  \bibnamefont {Beenakker}},\ }in\ \href
  {https://link.springer.com/chapter/10.1007/978-3-642-84818-6_22} {\emph
  {\bibinfo {booktitle} {Transport {Phenomena} in {Mesoscopic} {Systems}}}},\
  \bibinfo {series} {Springer {Series} in {Solid}-{State} {Sciences}}, Vol.\
  \bibinfo {volume} {109}\ (\bibinfo  {publisher} {Springer},\ \bibinfo
  {address} {Berlin},\ \bibinfo {year} {1992})\ pp.\ \bibinfo {pages}
  {235--253}\BibitemShut {NoStop}%
\bibitem [{\citenamefont {Fatemi}\ \emph {et~al.}(2022)\citenamefont {Fatemi},
  \citenamefont {Kurilovich}, \citenamefont {Hays}, \citenamefont {Bouman},
  \citenamefont {Connolly}, \citenamefont {Diamond}, \citenamefont {Frattini},
  \citenamefont {Kurilovich}, \citenamefont {Krogstrup}, \citenamefont
  {Nygård}, \citenamefont {Geresdi}, \citenamefont {Glazman},\ and\
  \citenamefont {Devoret}}]{fatemi_microwave_2022}%
  \BibitemOpen
  \bibfield  {author} {\bibinfo {author} {\bibfnamefont {V.}~\bibnamefont
  {Fatemi}}, \bibinfo {author} {\bibfnamefont {P.}~\bibnamefont {Kurilovich}},
  \bibinfo {author} {\bibfnamefont {M.}~\bibnamefont {Hays}}, \bibinfo {author}
  {\bibfnamefont {D.}~\bibnamefont {Bouman}}, \bibinfo {author} {\bibfnamefont
  {T.}~\bibnamefont {Connolly}}, \bibinfo {author} {\bibfnamefont
  {S.}~\bibnamefont {Diamond}}, \bibinfo {author} {\bibfnamefont
  {N.}~\bibnamefont {Frattini}}, \bibinfo {author} {\bibfnamefont
  {V.}~\bibnamefont {Kurilovich}}, \bibinfo {author} {\bibfnamefont
  {P.}~\bibnamefont {Krogstrup}}, \bibinfo {author} {\bibfnamefont
  {J.}~\bibnamefont {Nygård}}, \bibinfo {author} {\bibfnamefont
  {A.}~\bibnamefont {Geresdi}}, \bibinfo {author} {\bibfnamefont
  {L.}~\bibnamefont {Glazman}},\ and\ \bibinfo {author} {\bibfnamefont
  {M.}~\bibnamefont {Devoret}},\ }\href
  {https://doi.org/10.1103/PhysRevLett.129.227701} {\bibfield  {journal}
  {\bibinfo  {journal} {Physical Review Letters}\ }\textbf {\bibinfo {volume}
  {129}},\ \bibinfo {pages} {227701} (\bibinfo {year} {2022})}\BibitemShut
  {NoStop}%
\bibitem [{\citenamefont {Kruti}\ and\ \citenamefont
  {Riwar}(2024)}]{kruti_impact_2024}%
  \BibitemOpen
  \bibfield  {author} {\bibinfo {author} {\bibfnamefont {D.}~\bibnamefont
  {Kruti}}\ and\ \bibinfo {author} {\bibfnamefont {R.-P.}\ \bibnamefont
  {Riwar}},\ }\href {https://doi.org/10.48550/arXiv.2408.07035} {\bibinfo
  {title} {Impact of evanescent scattering modes and finite dispersion in
  superconducting junctions}} (\bibinfo {year} {2024}),\ \bibinfo {note}
  {arXiv:2408.07035 [cond-mat]}\BibitemShut {NoStop}%
\bibitem [{\citenamefont {Miano}\ \emph {et~al.}(2023)\citenamefont {Miano},
  \citenamefont {Joshi}, \citenamefont {Liu}, \citenamefont {Dai},
  \citenamefont {Parakh}, \citenamefont {Frunzio},\ and\ \citenamefont
  {Devoret}}]{miano_hamiltonian_2023}%
  \BibitemOpen
  \bibfield  {author} {\bibinfo {author} {\bibfnamefont {A.}~\bibnamefont
  {Miano}}, \bibinfo {author} {\bibfnamefont {V.}~\bibnamefont {Joshi}},
  \bibinfo {author} {\bibfnamefont {G.}~\bibnamefont {Liu}}, \bibinfo {author}
  {\bibfnamefont {W.}~\bibnamefont {Dai}}, \bibinfo {author} {\bibfnamefont
  {P.}~\bibnamefont {Parakh}}, \bibinfo {author} {\bibfnamefont
  {L.}~\bibnamefont {Frunzio}},\ and\ \bibinfo {author} {\bibfnamefont
  {M.}~\bibnamefont {Devoret}},\ }\href
  {https://doi.org/10.1103/PRXQuantum.4.030324} {\bibfield  {journal} {\bibinfo
   {journal} {PRX Quantum}\ }\textbf {\bibinfo {volume} {4}},\ \bibinfo {pages}
  {030324} (\bibinfo {year} {2023})}\BibitemShut {NoStop}%
\bibitem [{\citenamefont {Nazarov}\ and\ \citenamefont
  {Blanter}(2009)}]{nazarov_quantum_2009}%
  \BibitemOpen
  \bibfield  {author} {\bibinfo {author} {\bibfnamefont {Y.~V.}\ \bibnamefont
  {Nazarov}}\ and\ \bibinfo {author} {\bibfnamefont {Y.}~\bibnamefont
  {Blanter}},\ }\href@noop {} {\emph {\bibinfo {title} {Quantum {Transport}}}}\
  (\bibinfo  {publisher} {Cambridge University Press},\ \bibinfo {year}
  {2009})\BibitemShut {NoStop}%
\bibitem [{\citenamefont {Zazunov}\ \emph {et~al.}(2003)\citenamefont
  {Zazunov}, \citenamefont {Shumeiko}, \citenamefont {Bratus’}, \citenamefont
  {Lantz},\ and\ \citenamefont {Wendin}}]{zazunov_andreev_2003}%
  \BibitemOpen
  \bibfield  {author} {\bibinfo {author} {\bibfnamefont {A.}~\bibnamefont
  {Zazunov}}, \bibinfo {author} {\bibfnamefont {V.~S.}\ \bibnamefont
  {Shumeiko}}, \bibinfo {author} {\bibfnamefont {E.~N.}\ \bibnamefont
  {Bratus’}}, \bibinfo {author} {\bibfnamefont {J.}~\bibnamefont {Lantz}},\
  and\ \bibinfo {author} {\bibfnamefont {G.}~\bibnamefont {Wendin}},\ }\href
  {https://doi.org/10.1103/PhysRevLett.90.087003} {\bibfield  {journal}
  {\bibinfo  {journal} {Physical Review Letters}\ }\textbf {\bibinfo {volume}
  {90}},\ \bibinfo {pages} {087003} (\bibinfo {year} {2003})}\BibitemShut
  {NoStop}%
\bibitem [{\citenamefont {Pita-Vidal}\ \emph {et~al.}(2024)\citenamefont
  {Pita-Vidal}, \citenamefont {Wesdorp}, \citenamefont {Splitthoff},
  \citenamefont {Bargerbos}, \citenamefont {Liu}, \citenamefont {Kouwenhoven},\
  and\ \citenamefont {Andersen}}]{pita-vidal_strong_2024}%
  \BibitemOpen
  \bibfield  {author} {\bibinfo {author} {\bibfnamefont {M.}~\bibnamefont
  {Pita-Vidal}}, \bibinfo {author} {\bibfnamefont {J.~J.}\ \bibnamefont
  {Wesdorp}}, \bibinfo {author} {\bibfnamefont {L.~J.}\ \bibnamefont
  {Splitthoff}}, \bibinfo {author} {\bibfnamefont {A.}~\bibnamefont
  {Bargerbos}}, \bibinfo {author} {\bibfnamefont {Y.}~\bibnamefont {Liu}},
  \bibinfo {author} {\bibfnamefont {L.~P.}\ \bibnamefont {Kouwenhoven}},\ and\
  \bibinfo {author} {\bibfnamefont {C.~K.}\ \bibnamefont {Andersen}},\ }\href
  {https://doi.org/10.1038/s41567-024-02497-x} {\bibfield  {journal} {\bibinfo
  {journal} {Nature Physics}\ ,\ \bibinfo {pages} {1}} (\bibinfo {year}
  {2024})}\BibitemShut {NoStop}%
\bibitem [{\citenamefont {Vakhtel}\ \emph {et~al.}(2024)\citenamefont
  {Vakhtel}, \citenamefont {Kurilovich}, \citenamefont {Pita-Vidal},
  \citenamefont {Bargerbos}, \citenamefont {Fatemi},\ and\ \citenamefont {van
  Heck}}]{vakhtel_tunneling_2024}%
  \BibitemOpen
  \bibfield  {author} {\bibinfo {author} {\bibfnamefont {T.}~\bibnamefont
  {Vakhtel}}, \bibinfo {author} {\bibfnamefont {P.~D.}\ \bibnamefont
  {Kurilovich}}, \bibinfo {author} {\bibfnamefont {M.}~\bibnamefont
  {Pita-Vidal}}, \bibinfo {author} {\bibfnamefont {A.}~\bibnamefont
  {Bargerbos}}, \bibinfo {author} {\bibfnamefont {V.}~\bibnamefont {Fatemi}},\
  and\ \bibinfo {author} {\bibfnamefont {B.}~\bibnamefont {van Heck}},\ }\href
  {https://doi.org/10.1103/PhysRevB.110.045404} {\bibfield  {journal} {\bibinfo
   {journal} {Physical Review B}\ }\textbf {\bibinfo {volume} {110}},\ \bibinfo
  {pages} {045404} (\bibinfo {year} {2024})}\BibitemShut {NoStop}%
\bibitem [{\citenamefont {Cheung}\ \emph {et~al.}(2023)\citenamefont {Cheung},
  \citenamefont {Haller}, \citenamefont {Kononov}, \citenamefont {Ciaccia},
  \citenamefont {Ungerer}, \citenamefont {Kanne}, \citenamefont {Nygård},
  \citenamefont {Winkel}, \citenamefont {Reisinger}, \citenamefont {Pop},
  \citenamefont {Baumgartner},\ and\ \citenamefont
  {Schönenberger}}]{cheung_photon-mediated_2023}%
  \BibitemOpen
  \bibfield  {author} {\bibinfo {author} {\bibfnamefont {L.~Y.}\ \bibnamefont
  {Cheung}}, \bibinfo {author} {\bibfnamefont {R.}~\bibnamefont {Haller}},
  \bibinfo {author} {\bibfnamefont {A.}~\bibnamefont {Kononov}}, \bibinfo
  {author} {\bibfnamefont {C.}~\bibnamefont {Ciaccia}}, \bibinfo {author}
  {\bibfnamefont {J.~H.}\ \bibnamefont {Ungerer}}, \bibinfo {author}
  {\bibfnamefont {T.}~\bibnamefont {Kanne}}, \bibinfo {author} {\bibfnamefont
  {J.}~\bibnamefont {Nygård}}, \bibinfo {author} {\bibfnamefont
  {P.}~\bibnamefont {Winkel}}, \bibinfo {author} {\bibfnamefont
  {T.}~\bibnamefont {Reisinger}}, \bibinfo {author} {\bibfnamefont {I.~M.}\
  \bibnamefont {Pop}}, \bibinfo {author} {\bibfnamefont {A.}~\bibnamefont
  {Baumgartner}},\ and\ \bibinfo {author} {\bibfnamefont {C.}~\bibnamefont
  {Schönenberger}},\ }\href {https://doi.org/10.48550/arXiv.2310.15995}
  {\bibinfo {title} {Photon-mediated long range coupling of two {Andreev} level
  qubits}} (\bibinfo {year} {2023})\BibitemShut {NoStop}%
\bibitem [{\citenamefont {Averin}(1999)}]{averin_coulomb_1999}%
  \BibitemOpen
  \bibfield  {author} {\bibinfo {author} {\bibfnamefont {D.~V.}\ \bibnamefont
  {Averin}},\ }\href {https://doi.org/10.1103/PhysRevLett.82.3685} {\bibfield
  {journal} {\bibinfo  {journal} {Physical Review Letters}\ }\textbf {\bibinfo
  {volume} {82}},\ \bibinfo {pages} {3685} (\bibinfo {year}
  {1999})}\BibitemShut {NoStop}%
\bibitem [{\citenamefont {Bargerbos}\ \emph {et~al.}(2020)\citenamefont
  {Bargerbos}, \citenamefont {Uilhoorn}, \citenamefont {Yang}, \citenamefont
  {Krogstrup}, \citenamefont {Kouwenhoven}, \citenamefont {de~Lange},
  \citenamefont {van Heck},\ and\ \citenamefont
  {Kou}}]{bargerbos_observation_2020}%
  \BibitemOpen
  \bibfield  {author} {\bibinfo {author} {\bibfnamefont {A.}~\bibnamefont
  {Bargerbos}}, \bibinfo {author} {\bibfnamefont {W.}~\bibnamefont {Uilhoorn}},
  \bibinfo {author} {\bibfnamefont {C.-K.}\ \bibnamefont {Yang}}, \bibinfo
  {author} {\bibfnamefont {P.}~\bibnamefont {Krogstrup}}, \bibinfo {author}
  {\bibfnamefont {L.~P.}\ \bibnamefont {Kouwenhoven}}, \bibinfo {author}
  {\bibfnamefont {G.}~\bibnamefont {de~Lange}}, \bibinfo {author}
  {\bibfnamefont {B.}~\bibnamefont {van Heck}},\ and\ \bibinfo {author}
  {\bibfnamefont {A.}~\bibnamefont {Kou}},\ }\href
  {https://doi.org/10.1103/PhysRevLett.124.246802} {\bibfield  {journal}
  {\bibinfo  {journal} {Physical Review Letters}\ }\textbf {\bibinfo {volume}
  {124}},\ \bibinfo {pages} {246802} (\bibinfo {year} {2020})}\BibitemShut
  {NoStop}%
\bibitem [{\citenamefont {Kringhøj}\ \emph {et~al.}(2020)\citenamefont
  {Kringhøj}, \citenamefont {van Heck}, \citenamefont {Larsen}, \citenamefont
  {Erlandsson}, \citenamefont {Sabonis}, \citenamefont {Krogstrup},
  \citenamefont {Casparis}, \citenamefont {Petersson},\ and\ \citenamefont
  {Marcus}}]{kringhoj_suppressed_2020}%
  \BibitemOpen
  \bibfield  {author} {\bibinfo {author} {\bibfnamefont {A.}~\bibnamefont
  {Kringhøj}}, \bibinfo {author} {\bibfnamefont {B.}~\bibnamefont {van Heck}},
  \bibinfo {author} {\bibfnamefont {T.}~\bibnamefont {Larsen}}, \bibinfo
  {author} {\bibfnamefont {O.}~\bibnamefont {Erlandsson}}, \bibinfo {author}
  {\bibfnamefont {D.}~\bibnamefont {Sabonis}}, \bibinfo {author} {\bibfnamefont
  {P.}~\bibnamefont {Krogstrup}}, \bibinfo {author} {\bibfnamefont
  {L.}~\bibnamefont {Casparis}}, \bibinfo {author} {\bibfnamefont
  {K.}~\bibnamefont {Petersson}},\ and\ \bibinfo {author} {\bibfnamefont
  {C.}~\bibnamefont {Marcus}},\ }\href
  {https://doi.org/10.1103/PhysRevLett.124.246803} {\bibfield  {journal}
  {\bibinfo  {journal} {Physical Review Letters}\ }\textbf {\bibinfo {volume}
  {124}},\ \bibinfo {pages} {246803} (\bibinfo {year} {2020})}\BibitemShut
  {NoStop}%
\bibitem [{\citenamefont {Chtchelkatchev}\ and\ \citenamefont
  {Nazarov}(2003)}]{chtchelkatchev_andreev_2003}%
  \BibitemOpen
  \bibfield  {author} {\bibinfo {author} {\bibfnamefont {N.~M.}\ \bibnamefont
  {Chtchelkatchev}}\ and\ \bibinfo {author} {\bibfnamefont {Y.~V.}\
  \bibnamefont {Nazarov}},\ }\href
  {https://doi.org/10.1103/PhysRevLett.90.226806} {\bibfield  {journal}
  {\bibinfo  {journal} {Physical Review Letters}\ }\textbf {\bibinfo {volume}
  {90}},\ \bibinfo {pages} {226806} (\bibinfo {year} {2003})}\BibitemShut
  {NoStop}%
\bibitem [{\citenamefont {Park}\ and\ \citenamefont
  {Yeyati}(2017)}]{park_andreev_2017}%
  \BibitemOpen
  \bibfield  {author} {\bibinfo {author} {\bibfnamefont {S.}~\bibnamefont
  {Park}}\ and\ \bibinfo {author} {\bibfnamefont {A.~L.}\ \bibnamefont
  {Yeyati}},\ }\href {https://doi.org/10.1103/PhysRevB.96.125416} {\bibfield
  {journal} {\bibinfo  {journal} {Physical Review B}\ }\textbf {\bibinfo
  {volume} {96}},\ \bibinfo {pages} {125416} (\bibinfo {year}
  {2017})}\BibitemShut {NoStop}%
\bibitem [{\citenamefont {Tosi}\ \emph {et~al.}(2019)\citenamefont {Tosi},
  \citenamefont {Metzger}, \citenamefont {Goffman}, \citenamefont {Urbina},
  \citenamefont {Pothier}, \citenamefont {Park}, \citenamefont {Yeyati},
  \citenamefont {Nygård},\ and\ \citenamefont
  {Krogstrup}}]{tosi_spin-orbit_2019}%
  \BibitemOpen
  \bibfield  {author} {\bibinfo {author} {\bibfnamefont {L.}~\bibnamefont
  {Tosi}}, \bibinfo {author} {\bibfnamefont {C.}~\bibnamefont {Metzger}},
  \bibinfo {author} {\bibfnamefont {M.}~\bibnamefont {Goffman}}, \bibinfo
  {author} {\bibfnamefont {C.}~\bibnamefont {Urbina}}, \bibinfo {author}
  {\bibfnamefont {H.}~\bibnamefont {Pothier}}, \bibinfo {author} {\bibfnamefont
  {S.}~\bibnamefont {Park}}, \bibinfo {author} {\bibfnamefont {A.~L.}\
  \bibnamefont {Yeyati}}, \bibinfo {author} {\bibfnamefont {J.}~\bibnamefont
  {Nygård}},\ and\ \bibinfo {author} {\bibfnamefont {P.}~\bibnamefont
  {Krogstrup}},\ }\href {https://doi.org/10.1103/PhysRevX.9.011010} {\bibfield
  {journal} {\bibinfo  {journal} {Physical Review X}\ }\textbf {\bibinfo
  {volume} {9}},\ \bibinfo {pages} {011010} (\bibinfo {year}
  {2019})}\BibitemShut {NoStop}%
\bibitem [{\citenamefont {Hays}\ \emph {et~al.}(2021)\citenamefont {Hays},
  \citenamefont {Fatemi}, \citenamefont {Bouman}, \citenamefont {Cerrillo},
  \citenamefont {Diamond}, \citenamefont {Serniak}, \citenamefont {Connolly},
  \citenamefont {Krogstrup}, \citenamefont {Nygård}, \citenamefont {Yeyati},
  \citenamefont {Geresdi},\ and\ \citenamefont {Devoret}}]{hays_coherent_2021}%
  \BibitemOpen
  \bibfield  {author} {\bibinfo {author} {\bibfnamefont {M.}~\bibnamefont
  {Hays}}, \bibinfo {author} {\bibfnamefont {V.}~\bibnamefont {Fatemi}},
  \bibinfo {author} {\bibfnamefont {D.}~\bibnamefont {Bouman}}, \bibinfo
  {author} {\bibfnamefont {J.}~\bibnamefont {Cerrillo}}, \bibinfo {author}
  {\bibfnamefont {S.}~\bibnamefont {Diamond}}, \bibinfo {author} {\bibfnamefont
  {K.}~\bibnamefont {Serniak}}, \bibinfo {author} {\bibfnamefont
  {T.}~\bibnamefont {Connolly}}, \bibinfo {author} {\bibfnamefont
  {P.}~\bibnamefont {Krogstrup}}, \bibinfo {author} {\bibfnamefont
  {J.}~\bibnamefont {Nygård}}, \bibinfo {author} {\bibfnamefont {A.~L.}\
  \bibnamefont {Yeyati}}, \bibinfo {author} {\bibfnamefont {A.}~\bibnamefont
  {Geresdi}},\ and\ \bibinfo {author} {\bibfnamefont {M.~H.}\ \bibnamefont
  {Devoret}},\ }\href {https://doi.org/10.1126/science.abf0345} {\bibfield
  {journal} {\bibinfo  {journal} {Science}\ }\textbf {\bibinfo {volume}
  {373}},\ \bibinfo {pages} {430} (\bibinfo {year} {2021})}\BibitemShut
  {NoStop}%
\bibitem [{\citenamefont {Jarillo-Herrero}\ \emph {et~al.}(2006)\citenamefont
  {Jarillo-Herrero}, \citenamefont {van Dam},\ and\ \citenamefont
  {Kouwenhoven}}]{jarillo-herrero_quantum_2006}%
  \BibitemOpen
  \bibfield  {author} {\bibinfo {author} {\bibfnamefont {P.}~\bibnamefont
  {Jarillo-Herrero}}, \bibinfo {author} {\bibfnamefont {J.~A.}\ \bibnamefont
  {van Dam}},\ and\ \bibinfo {author} {\bibfnamefont {L.~P.}\ \bibnamefont
  {Kouwenhoven}},\ }\href {https://doi.org/10.1038/nature04550} {\bibfield
  {journal} {\bibinfo  {journal} {Nature}\ }\textbf {\bibinfo {volume} {439}},\
  \bibinfo {pages} {953} (\bibinfo {year} {2006})}\BibitemShut {NoStop}%
\bibitem [{\citenamefont {Heersche}\ \emph {et~al.}(2007)\citenamefont
  {Heersche}, \citenamefont {Jarillo-Herrero}, \citenamefont {Oostinga},
  \citenamefont {Vandersypen},\ and\ \citenamefont
  {Morpurgo}}]{heersche_bipolar_2007}%
  \BibitemOpen
  \bibfield  {author} {\bibinfo {author} {\bibfnamefont {H.~B.}\ \bibnamefont
  {Heersche}}, \bibinfo {author} {\bibfnamefont {P.}~\bibnamefont
  {Jarillo-Herrero}}, \bibinfo {author} {\bibfnamefont {J.~B.}\ \bibnamefont
  {Oostinga}}, \bibinfo {author} {\bibfnamefont {L.~M.~K.}\ \bibnamefont
  {Vandersypen}},\ and\ \bibinfo {author} {\bibfnamefont {A.~F.}\ \bibnamefont
  {Morpurgo}},\ }\href {https://doi.org/10.1038/nature05555} {\bibfield
  {journal} {\bibinfo  {journal} {Nature}\ }\textbf {\bibinfo {volume} {446}},\
  \bibinfo {pages} {56} (\bibinfo {year} {2007})}\BibitemShut {NoStop}%
\bibitem [{\citenamefont {Larsen}\ \emph
  {et~al.}(2015{\natexlab{a}})\citenamefont {Larsen}, \citenamefont
  {Petersson}, \citenamefont {Kuemmeth}, \citenamefont {Jespersen},
  \citenamefont {Krogstrup}, \citenamefont {Nygård},\ and\ \citenamefont
  {Marcus}}]{larsen_semiconductor-nanowire-based_2015}%
  \BibitemOpen
  \bibfield  {author} {\bibinfo {author} {\bibfnamefont {T.}~\bibnamefont
  {Larsen}}, \bibinfo {author} {\bibfnamefont {K.}~\bibnamefont {Petersson}},
  \bibinfo {author} {\bibfnamefont {F.}~\bibnamefont {Kuemmeth}}, \bibinfo
  {author} {\bibfnamefont {T.}~\bibnamefont {Jespersen}}, \bibinfo {author}
  {\bibfnamefont {P.}~\bibnamefont {Krogstrup}}, \bibinfo {author}
  {\bibfnamefont {J.}~\bibnamefont {Nygård}},\ and\ \bibinfo {author}
  {\bibfnamefont {C.}~\bibnamefont {Marcus}},\ }\href
  {https://doi.org/10.1103/PhysRevLett.115.127001} {\bibfield  {journal}
  {\bibinfo  {journal} {Physical Review Letters}\ }\textbf {\bibinfo {volume}
  {115}},\ \bibinfo {pages} {127001} (\bibinfo {year}
  {2015}{\natexlab{a}})}\BibitemShut {NoStop}%
\bibitem [{\citenamefont {Heedt}\ \emph {et~al.}(2021)\citenamefont {Heedt},
  \citenamefont {{Quintero-P{\'e}rez}}, \citenamefont {Borsoi}, \citenamefont
  {Fursina}, \citenamefont {{van Loo}}, \citenamefont {Mazur}, \citenamefont
  {Nowak}, \citenamefont {Ammerlaan}, \citenamefont {Li}, \citenamefont
  {Korneychuk}, \citenamefont {Shen}, \citenamefont {{van de Poll}},
  \citenamefont {Badawy}, \citenamefont {Gazibegovic}, \citenamefont {{de
  Jong}}, \citenamefont {Aseev}, \citenamefont {{van Hoogdalem}}, \citenamefont
  {Bakkers},\ and\ \citenamefont {Kouwenhoven}}]{heedt2021Shadowwall}%
  \BibitemOpen
  \bibfield  {author} {\bibinfo {author} {\bibfnamefont {S.}~\bibnamefont
  {Heedt}}, \bibinfo {author} {\bibfnamefont {M.}~\bibnamefont
  {{Quintero-P{\'e}rez}}}, \bibinfo {author} {\bibfnamefont {F.}~\bibnamefont
  {Borsoi}}, \bibinfo {author} {\bibfnamefont {A.}~\bibnamefont {Fursina}},
  \bibinfo {author} {\bibfnamefont {N.}~\bibnamefont {{van Loo}}}, \bibinfo
  {author} {\bibfnamefont {G.~P.}\ \bibnamefont {Mazur}}, \bibinfo {author}
  {\bibfnamefont {M.~P.}\ \bibnamefont {Nowak}}, \bibinfo {author}
  {\bibfnamefont {M.}~\bibnamefont {Ammerlaan}}, \bibinfo {author}
  {\bibfnamefont {K.}~\bibnamefont {Li}}, \bibinfo {author} {\bibfnamefont
  {S.}~\bibnamefont {Korneychuk}}, \bibinfo {author} {\bibfnamefont
  {J.}~\bibnamefont {Shen}}, \bibinfo {author} {\bibfnamefont {M.~A.~Y.}\
  \bibnamefont {{van de Poll}}}, \bibinfo {author} {\bibfnamefont
  {G.}~\bibnamefont {Badawy}}, \bibinfo {author} {\bibfnamefont
  {S.}~\bibnamefont {Gazibegovic}}, \bibinfo {author} {\bibfnamefont
  {N.}~\bibnamefont {{de Jong}}}, \bibinfo {author} {\bibfnamefont
  {P.}~\bibnamefont {Aseev}}, \bibinfo {author} {\bibfnamefont
  {K.}~\bibnamefont {{van Hoogdalem}}}, \bibinfo {author} {\bibfnamefont {E.~P.
  A.~M.}\ \bibnamefont {Bakkers}},\ and\ \bibinfo {author} {\bibfnamefont
  {L.~P.}\ \bibnamefont {Kouwenhoven}},\ }\href
  {https://doi.org/10.1038/s41467-021-25100-w} {\bibfield  {journal} {\bibinfo
  {journal} {Nat. Commun}\ }\textbf {\bibinfo {volume} {12}},\ \bibinfo {pages}
  {4914} (\bibinfo {year} {2021})}\BibitemShut {NoStop}%
\bibitem [{\citenamefont {{van Loo}}\ \emph {et~al.}(2023)\citenamefont {{van
  Loo}}, \citenamefont {Mazur}, \citenamefont {Dvir}, \citenamefont {Wang},
  \citenamefont {Dekker}, \citenamefont {Wang}, \citenamefont {Lemang},
  \citenamefont {Sfiligoj}, \citenamefont {Bordin}, \citenamefont {{van
  Driel}}, \citenamefont {Badawy}, \citenamefont {Gazibegovic}, \citenamefont
  {Bakkers},\ and\ \citenamefont {Kouwenhoven}}]{vanloo2023Electrostatic}%
  \BibitemOpen
  \bibfield  {author} {\bibinfo {author} {\bibfnamefont {N.}~\bibnamefont {{van
  Loo}}}, \bibinfo {author} {\bibfnamefont {G.~P.}\ \bibnamefont {Mazur}},
  \bibinfo {author} {\bibfnamefont {T.}~\bibnamefont {Dvir}}, \bibinfo {author}
  {\bibfnamefont {G.}~\bibnamefont {Wang}}, \bibinfo {author} {\bibfnamefont
  {R.~C.}\ \bibnamefont {Dekker}}, \bibinfo {author} {\bibfnamefont {J.-Y.}\
  \bibnamefont {Wang}}, \bibinfo {author} {\bibfnamefont {M.}~\bibnamefont
  {Lemang}}, \bibinfo {author} {\bibfnamefont {C.}~\bibnamefont {Sfiligoj}},
  \bibinfo {author} {\bibfnamefont {A.}~\bibnamefont {Bordin}}, \bibinfo
  {author} {\bibfnamefont {D.}~\bibnamefont {{van Driel}}}, \bibinfo {author}
  {\bibfnamefont {G.}~\bibnamefont {Badawy}}, \bibinfo {author} {\bibfnamefont
  {S.}~\bibnamefont {Gazibegovic}}, \bibinfo {author} {\bibfnamefont {E.~P.
  A.~M.}\ \bibnamefont {Bakkers}},\ and\ \bibinfo {author} {\bibfnamefont
  {L.~P.}\ \bibnamefont {Kouwenhoven}},\ }\href
  {https://doi.org/10.1038/s41467-023-39044-w} {\bibfield  {journal} {\bibinfo
  {journal} {Nat. Commun}\ }\textbf {\bibinfo {volume} {14}},\ \bibinfo {pages}
  {3325} (\bibinfo {year} {2023})}\BibitemShut {NoStop}%
\bibitem [{\citenamefont {Kjaergaard}\ \emph {et~al.}(2020)\citenamefont
  {Kjaergaard}, \citenamefont {Schwartz}, \citenamefont {Braumüller},
  \citenamefont {Krantz}, \citenamefont {Wang}, \citenamefont {Gustavsson},\
  and\ \citenamefont {Oliver}}]{kjaergaard_superconducting_2020}%
  \BibitemOpen
  \bibfield  {author} {\bibinfo {author} {\bibfnamefont {M.}~\bibnamefont
  {Kjaergaard}}, \bibinfo {author} {\bibfnamefont {M.~E.}\ \bibnamefont
  {Schwartz}}, \bibinfo {author} {\bibfnamefont {J.}~\bibnamefont
  {Braumüller}}, \bibinfo {author} {\bibfnamefont {P.}~\bibnamefont {Krantz}},
  \bibinfo {author} {\bibfnamefont {J.~I.-J.}\ \bibnamefont {Wang}}, \bibinfo
  {author} {\bibfnamefont {S.}~\bibnamefont {Gustavsson}},\ and\ \bibinfo
  {author} {\bibfnamefont {W.~D.}\ \bibnamefont {Oliver}},\ }\href
  {https://doi.org/10.1146/annurev-conmatphys-031119-050605} {\bibfield
  {journal} {\bibinfo  {journal} {Annual Review of Condensed Matter Physics}\
  }\textbf {\bibinfo {volume} {11}},\ \bibinfo {pages} {369} (\bibinfo {year}
  {2020})}\BibitemShut {NoStop}%
\bibitem [{\citenamefont {Blais}\ \emph {et~al.}(2021)\citenamefont {Blais},
  \citenamefont {Grimsmo}, \citenamefont {Girvin},\ and\ \citenamefont
  {Wallraff}}]{blais_circuit_2021}%
  \BibitemOpen
  \bibfield  {author} {\bibinfo {author} {\bibfnamefont {A.}~\bibnamefont
  {Blais}}, \bibinfo {author} {\bibfnamefont {A.~L.}\ \bibnamefont {Grimsmo}},
  \bibinfo {author} {\bibfnamefont {S.~M.}\ \bibnamefont {Girvin}},\ and\
  \bibinfo {author} {\bibfnamefont {A.}~\bibnamefont {Wallraff}},\ }\href
  {https://doi.org/10.1103/RevModPhys.93.025005} {\bibfield  {journal}
  {\bibinfo  {journal} {Reviews of Modern Physics}\ }\textbf {\bibinfo {volume}
  {93}},\ \bibinfo {pages} {025005} (\bibinfo {year} {2021})}\BibitemShut
  {NoStop}%
\bibitem [{\citenamefont {Banszerus}\ \emph
  {et~al.}(2024{\natexlab{b}})\citenamefont {Banszerus}, \citenamefont
  {Marshall}, \citenamefont {Andersson}, \citenamefont {Lindemann},
  \citenamefont {Manfra}, \citenamefont {Marcus},\ and\ \citenamefont
  {Vaitiekėnas}}]{banszerus_voltage-controlled_2024}%
  \BibitemOpen
  \bibfield  {author} {\bibinfo {author} {\bibfnamefont {L.}~\bibnamefont
  {Banszerus}}, \bibinfo {author} {\bibfnamefont {W.}~\bibnamefont {Marshall}},
  \bibinfo {author} {\bibfnamefont {C.~W.}\ \bibnamefont {Andersson}}, \bibinfo
  {author} {\bibfnamefont {T.}~\bibnamefont {Lindemann}}, \bibinfo {author}
  {\bibfnamefont {M.~J.}\ \bibnamefont {Manfra}}, \bibinfo {author}
  {\bibfnamefont {C.~M.}\ \bibnamefont {Marcus}},\ and\ \bibinfo {author}
  {\bibfnamefont {S.}~\bibnamefont {Vaitiekėnas}},\ }\href
  {https://doi.org/10.48550/arXiv.2402.11603} {\bibinfo {title}
  {Voltage-controlled synthesis of higher harmonics in hybrid {Josephson}
  junction circuits}} (\bibinfo {year} {2024}{\natexlab{b}}),\ \bibinfo {note}
  {arXiv:2402.11603 [cond-mat]}\BibitemShut {NoStop}%
\bibitem [{\citenamefont {Larsen}\ \emph
  {et~al.}(2015{\natexlab{b}})\citenamefont {Larsen}, \citenamefont
  {Petersson}, \citenamefont {Kuemmeth}, \citenamefont {Jespersen},
  \citenamefont {Krogstrup}, \citenamefont {Nyg{\aa}rd},\ and\ \citenamefont
  {Marcus}}]{larsen2015SemiconductorNanowireBased}%
  \BibitemOpen
  \bibfield  {author} {\bibinfo {author} {\bibfnamefont {T.~W.}\ \bibnamefont
  {Larsen}}, \bibinfo {author} {\bibfnamefont {K.~D.}\ \bibnamefont
  {Petersson}}, \bibinfo {author} {\bibfnamefont {F.}~\bibnamefont {Kuemmeth}},
  \bibinfo {author} {\bibfnamefont {T.~S.}\ \bibnamefont {Jespersen}}, \bibinfo
  {author} {\bibfnamefont {P.}~\bibnamefont {Krogstrup}}, \bibinfo {author}
  {\bibfnamefont {J.}~\bibnamefont {Nyg{\aa}rd}},\ and\ \bibinfo {author}
  {\bibfnamefont {C.~M.}\ \bibnamefont {Marcus}},\ }\href
  {https://doi.org/10.1103/PhysRevLett.115.127001} {\bibfield  {journal}
  {\bibinfo  {journal} {Phys. Rev. Lett.}\ }\textbf {\bibinfo {volume} {115}},\
  \bibinfo {pages} {127001} (\bibinfo {year} {2015}{\natexlab{b}})}\BibitemShut
  {NoStop}%
\bibitem [{\citenamefont {Casparis}\ \emph {et~al.}(2018)\citenamefont
  {Casparis}, \citenamefont {Connolly}, \citenamefont {Kjaergaard},
  \citenamefont {Pearson}, \citenamefont {Kringh{\o}j}, \citenamefont {Larsen},
  \citenamefont {Kuemmeth}, \citenamefont {Wang}, \citenamefont {Thomas},
  \citenamefont {Gronin}, \citenamefont {Gardner}, \citenamefont {Manfra},
  \citenamefont {Marcus},\ and\ \citenamefont
  {Petersson}}]{casparis2018Superconducting}%
  \BibitemOpen
  \bibfield  {author} {\bibinfo {author} {\bibfnamefont {L.}~\bibnamefont
  {Casparis}}, \bibinfo {author} {\bibfnamefont {M.~R.}\ \bibnamefont
  {Connolly}}, \bibinfo {author} {\bibfnamefont {M.}~\bibnamefont
  {Kjaergaard}}, \bibinfo {author} {\bibfnamefont {N.~J.}\ \bibnamefont
  {Pearson}}, \bibinfo {author} {\bibfnamefont {A.}~\bibnamefont
  {Kringh{\o}j}}, \bibinfo {author} {\bibfnamefont {T.~W.}\ \bibnamefont
  {Larsen}}, \bibinfo {author} {\bibfnamefont {F.}~\bibnamefont {Kuemmeth}},
  \bibinfo {author} {\bibfnamefont {T.}~\bibnamefont {Wang}}, \bibinfo {author}
  {\bibfnamefont {C.}~\bibnamefont {Thomas}}, \bibinfo {author} {\bibfnamefont
  {S.}~\bibnamefont {Gronin}}, \bibinfo {author} {\bibfnamefont {G.~C.}\
  \bibnamefont {Gardner}}, \bibinfo {author} {\bibfnamefont {M.~J.}\
  \bibnamefont {Manfra}}, \bibinfo {author} {\bibfnamefont {C.~M.}\
  \bibnamefont {Marcus}},\ and\ \bibinfo {author} {\bibfnamefont {K.~D.}\
  \bibnamefont {Petersson}},\ }\href
  {https://doi.org/10.1038/s41565-018-0207-y} {\bibfield  {journal} {\bibinfo
  {journal} {Nat. Nanotechnol.}\ }\textbf {\bibinfo {volume} {13}},\ \bibinfo
  {pages} {915} (\bibinfo {year} {2018})}\BibitemShut {NoStop}%
\bibitem [{\citenamefont {Casparis}\ \emph {et~al.}(2019)\citenamefont
  {Casparis}, \citenamefont {Pearson}, \citenamefont {Kringh{\o}j},
  \citenamefont {Larsen}, \citenamefont {Kuemmeth}, \citenamefont {Nyg{\aa}rd},
  \citenamefont {Krogstrup}, \citenamefont {Petersson},\ and\ \citenamefont
  {Marcus}}]{casparis2019Voltagecontrolled}%
  \BibitemOpen
  \bibfield  {author} {\bibinfo {author} {\bibfnamefont {L.}~\bibnamefont
  {Casparis}}, \bibinfo {author} {\bibfnamefont {N.~J.}\ \bibnamefont
  {Pearson}}, \bibinfo {author} {\bibfnamefont {A.}~\bibnamefont
  {Kringh{\o}j}}, \bibinfo {author} {\bibfnamefont {T.~W.}\ \bibnamefont
  {Larsen}}, \bibinfo {author} {\bibfnamefont {F.}~\bibnamefont {Kuemmeth}},
  \bibinfo {author} {\bibfnamefont {J.}~\bibnamefont {Nyg{\aa}rd}}, \bibinfo
  {author} {\bibfnamefont {P.}~\bibnamefont {Krogstrup}}, \bibinfo {author}
  {\bibfnamefont {K.~D.}\ \bibnamefont {Petersson}},\ and\ \bibinfo {author}
  {\bibfnamefont {C.~M.}\ \bibnamefont {Marcus}},\ }\href
  {https://doi.org/10.1103/PhysRevB.99.085434} {\bibfield  {journal} {\bibinfo
  {journal} {Phys. Rev. B}\ }\textbf {\bibinfo {volume} {99}},\ \bibinfo
  {pages} {085434} (\bibinfo {year} {2019})}\BibitemShut {NoStop}%
\bibitem [{\citenamefont {Sardashti}\ \emph {et~al.}(2020)\citenamefont
  {Sardashti}, \citenamefont {Dartiailh}, \citenamefont {Yuan}, \citenamefont
  {Hart}, \citenamefont {Gumann},\ and\ \citenamefont
  {Shabani}}]{sardashti2020VoltageTunable}%
  \BibitemOpen
  \bibfield  {author} {\bibinfo {author} {\bibfnamefont {K.}~\bibnamefont
  {Sardashti}}, \bibinfo {author} {\bibfnamefont {M.~C.}\ \bibnamefont
  {Dartiailh}}, \bibinfo {author} {\bibfnamefont {J.}~\bibnamefont {Yuan}},
  \bibinfo {author} {\bibfnamefont {S.}~\bibnamefont {Hart}}, \bibinfo {author}
  {\bibfnamefont {P.}~\bibnamefont {Gumann}},\ and\ \bibinfo {author}
  {\bibfnamefont {J.}~\bibnamefont {Shabani}},\ }\href
  {https://doi.org/10.1109/TQE.2020.3034553} {\bibfield  {journal} {\bibinfo
  {journal} {IEEE Transactions on Quantum Engineering}\ }\textbf {\bibinfo
  {volume} {1}},\ \bibinfo {pages} {1} (\bibinfo {year} {2020})}\BibitemShut
  {NoStop}%
\bibitem [{\citenamefont {Strickland}\ \emph {et~al.}(2023)\citenamefont
  {Strickland}, \citenamefont {Elfeky}, \citenamefont {Yuan}, \citenamefont
  {Schiela}, \citenamefont {Yu}, \citenamefont {Langone}, \citenamefont
  {Vavilov}, \citenamefont {Manucharyan},\ and\ \citenamefont
  {Shabani}}]{strickland2023Superconducting}%
  \BibitemOpen
  \bibfield  {author} {\bibinfo {author} {\bibfnamefont {W.~M.}\ \bibnamefont
  {Strickland}}, \bibinfo {author} {\bibfnamefont {B.~H.}\ \bibnamefont
  {Elfeky}}, \bibinfo {author} {\bibfnamefont {J.~O.}\ \bibnamefont {Yuan}},
  \bibinfo {author} {\bibfnamefont {W.~F.}\ \bibnamefont {Schiela}}, \bibinfo
  {author} {\bibfnamefont {P.}~\bibnamefont {Yu}}, \bibinfo {author}
  {\bibfnamefont {D.}~\bibnamefont {Langone}}, \bibinfo {author} {\bibfnamefont
  {M.~G.}\ \bibnamefont {Vavilov}}, \bibinfo {author} {\bibfnamefont {V.~E.}\
  \bibnamefont {Manucharyan}},\ and\ \bibinfo {author} {\bibfnamefont
  {J.}~\bibnamefont {Shabani}},\ }\href
  {https://doi.org/10.1103/PhysRevApplied.19.034021} {\bibfield  {journal}
  {\bibinfo  {journal} {Phys. Rev. Appl.}\ }\textbf {\bibinfo {volume} {19}},\
  \bibinfo {pages} {034021} (\bibinfo {year} {2023})}\BibitemShut {NoStop}%
\bibitem [{\citenamefont {Butseraen}\ \emph {et~al.}(2022)\citenamefont
  {Butseraen}, \citenamefont {Ranadive}, \citenamefont {Aparicio},
  \citenamefont {Rafsanjani~Amin}, \citenamefont {Juyal}, \citenamefont
  {Esposito}, \citenamefont {Watanabe}, \citenamefont {Taniguchi},
  \citenamefont {Roch}, \citenamefont {Lefloch},\ and\ \citenamefont
  {Renard}}]{butseraen2022gatetunable}%
  \BibitemOpen
  \bibfield  {author} {\bibinfo {author} {\bibfnamefont {G.}~\bibnamefont
  {Butseraen}}, \bibinfo {author} {\bibfnamefont {A.}~\bibnamefont {Ranadive}},
  \bibinfo {author} {\bibfnamefont {N.}~\bibnamefont {Aparicio}}, \bibinfo
  {author} {\bibfnamefont {K.}~\bibnamefont {Rafsanjani~Amin}}, \bibinfo
  {author} {\bibfnamefont {A.}~\bibnamefont {Juyal}}, \bibinfo {author}
  {\bibfnamefont {M.}~\bibnamefont {Esposito}}, \bibinfo {author}
  {\bibfnamefont {K.}~\bibnamefont {Watanabe}}, \bibinfo {author}
  {\bibfnamefont {T.}~\bibnamefont {Taniguchi}}, \bibinfo {author}
  {\bibfnamefont {N.}~\bibnamefont {Roch}}, \bibinfo {author} {\bibfnamefont
  {F.}~\bibnamefont {Lefloch}},\ and\ \bibinfo {author} {\bibfnamefont
  {J.}~\bibnamefont {Renard}},\ }\href
  {https://doi.org/10.1038/s41565-022-01235-9} {\bibfield  {journal} {\bibinfo
  {journal} {Nat. Nanotechnol.}\ }\textbf {\bibinfo {volume} {17}},\ \bibinfo
  {pages} {1153} (\bibinfo {year} {2022})}\BibitemShut {NoStop}%
\bibitem [{\citenamefont {Phan}\ \emph {et~al.}(2023)\citenamefont {Phan},
  \citenamefont {{Falthansl-Scheinecker}}, \citenamefont {Mishra},
  \citenamefont {Strickland}, \citenamefont {Langone}, \citenamefont
  {Shabani},\ and\ \citenamefont {Higginbotham}}]{phan2023GateTunable}%
  \BibitemOpen
  \bibfield  {author} {\bibinfo {author} {\bibfnamefont {D.}~\bibnamefont
  {Phan}}, \bibinfo {author} {\bibfnamefont {P.}~\bibnamefont
  {{Falthansl-Scheinecker}}}, \bibinfo {author} {\bibfnamefont
  {U.}~\bibnamefont {Mishra}}, \bibinfo {author} {\bibfnamefont
  {W.}~\bibnamefont {Strickland}}, \bibinfo {author} {\bibfnamefont
  {D.}~\bibnamefont {Langone}}, \bibinfo {author} {\bibfnamefont
  {J.}~\bibnamefont {Shabani}},\ and\ \bibinfo {author} {\bibfnamefont
  {A.}~\bibnamefont {Higginbotham}},\ }\href
  {https://doi.org/10.1103/PhysRevApplied.19.064032} {\bibfield  {journal}
  {\bibinfo  {journal} {Phys. Rev. Appl.}\ }\textbf {\bibinfo {volume} {19}},\
  \bibinfo {pages} {064032} (\bibinfo {year} {2023})}\BibitemShut {NoStop}%
\bibitem [{\citenamefont {Sarkar}\ \emph {et~al.}(2022)\citenamefont {Sarkar},
  \citenamefont {Salunkhe}, \citenamefont {Mandal}, \citenamefont {Ghatak},
  \citenamefont {Marchawala}, \citenamefont {Das}, \citenamefont {Watanabe},
  \citenamefont {Taniguchi}, \citenamefont {Vijay},\ and\ \citenamefont
  {Deshmukh}}]{sarkar2022Quantumnoiselimited}%
  \BibitemOpen
  \bibfield  {author} {\bibinfo {author} {\bibfnamefont {J.}~\bibnamefont
  {Sarkar}}, \bibinfo {author} {\bibfnamefont {K.~V.}\ \bibnamefont
  {Salunkhe}}, \bibinfo {author} {\bibfnamefont {S.}~\bibnamefont {Mandal}},
  \bibinfo {author} {\bibfnamefont {S.}~\bibnamefont {Ghatak}}, \bibinfo
  {author} {\bibfnamefont {A.~H.}\ \bibnamefont {Marchawala}}, \bibinfo
  {author} {\bibfnamefont {I.}~\bibnamefont {Das}}, \bibinfo {author}
  {\bibfnamefont {K.}~\bibnamefont {Watanabe}}, \bibinfo {author}
  {\bibfnamefont {T.}~\bibnamefont {Taniguchi}}, \bibinfo {author}
  {\bibfnamefont {R.}~\bibnamefont {Vijay}},\ and\ \bibinfo {author}
  {\bibfnamefont {M.~M.}\ \bibnamefont {Deshmukh}},\ }\href
  {https://doi.org/10.1038/s41565-022-01223-z} {\bibfield  {journal} {\bibinfo
  {journal} {Nat. Nanotechnol.}\ }\textbf {\bibinfo {volume} {17}},\ \bibinfo
  {pages} {1147} (\bibinfo {year} {2022})}\BibitemShut {NoStop}%
\bibitem [{\citenamefont {Hao}\ \emph {et~al.}(2024)\citenamefont {Hao},
  \citenamefont {Shaw}, \citenamefont {Hatefipour}, \citenamefont {Strickland},
  \citenamefont {Elfeky}, \citenamefont {Langone}, \citenamefont {Shabani},\
  and\ \citenamefont {Shankar}}]{hao_kerr_2024}%
  \BibitemOpen
  \bibfield  {author} {\bibinfo {author} {\bibfnamefont {Z.}~\bibnamefont
  {Hao}}, \bibinfo {author} {\bibfnamefont {T.}~\bibnamefont {Shaw}}, \bibinfo
  {author} {\bibfnamefont {M.}~\bibnamefont {Hatefipour}}, \bibinfo {author}
  {\bibfnamefont {W.~M.}\ \bibnamefont {Strickland}}, \bibinfo {author}
  {\bibfnamefont {B.~H.}\ \bibnamefont {Elfeky}}, \bibinfo {author}
  {\bibfnamefont {D.}~\bibnamefont {Langone}}, \bibinfo {author} {\bibfnamefont
  {J.}~\bibnamefont {Shabani}},\ and\ \bibinfo {author} {\bibfnamefont
  {S.}~\bibnamefont {Shankar}},\ }\href {https://doi.org/10.1063/5.0205053}
  {\bibfield  {journal} {\bibinfo  {journal} {Applied Physics Letters}\
  }\textbf {\bibinfo {volume} {124}},\ \bibinfo {pages} {254003} (\bibinfo
  {year} {2024})}\BibitemShut {NoStop}%
\bibitem [{\citenamefont {Leroux}\ \emph {et~al.}(2022)\citenamefont {Leroux},
  \citenamefont {{Parra-Rodriguez}}, \citenamefont {Shillito}, \citenamefont
  {Di~Paolo}, \citenamefont {Oliver}, \citenamefont {Marcus}, \citenamefont
  {Kjaergaard}, \citenamefont {Gyenis},\ and\ \citenamefont
  {Blais}}]{leroux2022Nonreciprocal}%
  \BibitemOpen
  \bibfield  {author} {\bibinfo {author} {\bibfnamefont {C.}~\bibnamefont
  {Leroux}}, \bibinfo {author} {\bibfnamefont {A.}~\bibnamefont
  {{Parra-Rodriguez}}}, \bibinfo {author} {\bibfnamefont {R.}~\bibnamefont
  {Shillito}}, \bibinfo {author} {\bibfnamefont {A.}~\bibnamefont {Di~Paolo}},
  \bibinfo {author} {\bibfnamefont {W.~D.}\ \bibnamefont {Oliver}}, \bibinfo
  {author} {\bibfnamefont {C.~M.}\ \bibnamefont {Marcus}}, \bibinfo {author}
  {\bibfnamefont {M.}~\bibnamefont {Kjaergaard}}, \bibinfo {author}
  {\bibfnamefont {A.}~\bibnamefont {Gyenis}},\ and\ \bibinfo {author}
  {\bibfnamefont {A.}~\bibnamefont {Blais}},\ }\bibfield  {journal} {\bibinfo
  {journal} {arXiv preprint arXiv:2209.06194}\ }\href
  {https://doi.org/10.48550/arXiv.2209.06194} {10.48550/arXiv.2209.06194}
  (\bibinfo {year} {2022})\BibitemShut {NoStop}%
\bibitem [{\citenamefont {Kreikebaum}\ \emph {et~al.}(2020)\citenamefont
  {Kreikebaum}, \citenamefont {O'Brien}, \citenamefont {Morvan},\ and\
  \citenamefont {Siddiqi}}]{kreikebaum_improving_2020}%
  \BibitemOpen
  \bibfield  {author} {\bibinfo {author} {\bibfnamefont {J.~M.}\ \bibnamefont
  {Kreikebaum}}, \bibinfo {author} {\bibfnamefont {K.~P.}\ \bibnamefont
  {O'Brien}}, \bibinfo {author} {\bibfnamefont {A.}~\bibnamefont {Morvan}},\
  and\ \bibinfo {author} {\bibfnamefont {I.}~\bibnamefont {Siddiqi}},\ }\href
  {https://doi.org/10.1088/1361-6668/ab8617} {\bibfield  {journal} {\bibinfo
  {journal} {Superconductor Science and Technology}\ }\textbf {\bibinfo
  {volume} {33}},\ \bibinfo {pages} {06LT02} (\bibinfo {year}
  {2020})}\BibitemShut {NoStop}%
\bibitem [{\citenamefont {Takahashi}\ \emph {et~al.}(2022)\citenamefont
  {Takahashi}, \citenamefont {Kouma}, \citenamefont {Doi}, \citenamefont
  {Sato}, \citenamefont {Tamate},\ and\ \citenamefont
  {Nakamura}}]{takahashi_uniformity_2022}%
  \BibitemOpen
  \bibfield  {author} {\bibinfo {author} {\bibfnamefont {T.}~\bibnamefont
  {Takahashi}}, \bibinfo {author} {\bibfnamefont {N.}~\bibnamefont {Kouma}},
  \bibinfo {author} {\bibfnamefont {Y.}~\bibnamefont {Doi}}, \bibinfo {author}
  {\bibfnamefont {S.}~\bibnamefont {Sato}}, \bibinfo {author} {\bibfnamefont
  {S.}~\bibnamefont {Tamate}},\ and\ \bibinfo {author} {\bibfnamefont
  {Y.}~\bibnamefont {Nakamura}},\ }\href
  {https://doi.org/10.35848/1347-4065/aca256} {\bibfield  {journal} {\bibinfo
  {journal} {Japanese Journal of Applied Physics}\ }\textbf {\bibinfo {volume}
  {62}},\ \bibinfo {pages} {SC1002} (\bibinfo {year} {2022})}\BibitemShut
  {NoStop}%
\bibitem [{\citenamefont {Zheng}\ \emph {et~al.}(2023)\citenamefont {Zheng},
  \citenamefont {Li}, \citenamefont {Ding}, \citenamefont {Xiong},
  \citenamefont {Feng},\ and\ \citenamefont {Yang}}]{zheng_fabrication_2023}%
  \BibitemOpen
  \bibfield  {author} {\bibinfo {author} {\bibfnamefont {Y.}~\bibnamefont
  {Zheng}}, \bibinfo {author} {\bibfnamefont {S.}~\bibnamefont {Li}}, \bibinfo
  {author} {\bibfnamefont {Z.}~\bibnamefont {Ding}}, \bibinfo {author}
  {\bibfnamefont {K.}~\bibnamefont {Xiong}}, \bibinfo {author} {\bibfnamefont
  {J.}~\bibnamefont {Feng}},\ and\ \bibinfo {author} {\bibfnamefont
  {H.}~\bibnamefont {Yang}},\ }\href
  {https://doi.org/10.1038/s41598-023-39052-2} {\bibfield  {journal} {\bibinfo
  {journal} {Scientific Reports}\ }\textbf {\bibinfo {volume} {13}},\ \bibinfo
  {pages} {11874} (\bibinfo {year} {2023})}\BibitemShut {NoStop}%
\bibitem [{\citenamefont {Krogstrup}\ \emph {et~al.}(2015)\citenamefont
  {Krogstrup}, \citenamefont {Ziino}, \citenamefont {Chang}, \citenamefont
  {Albrecht}, \citenamefont {Madsen}, \citenamefont {Johnson}, \citenamefont
  {Nygård}, \citenamefont {Marcus},\ and\ \citenamefont
  {Jespersen}}]{krogstrup_epitaxy_2015}%
  \BibitemOpen
  \bibfield  {author} {\bibinfo {author} {\bibfnamefont {P.}~\bibnamefont
  {Krogstrup}}, \bibinfo {author} {\bibfnamefont {N.~L.~B.}\ \bibnamefont
  {Ziino}}, \bibinfo {author} {\bibfnamefont {W.}~\bibnamefont {Chang}},
  \bibinfo {author} {\bibfnamefont {S.~M.}\ \bibnamefont {Albrecht}}, \bibinfo
  {author} {\bibfnamefont {M.~H.}\ \bibnamefont {Madsen}}, \bibinfo {author}
  {\bibfnamefont {E.}~\bibnamefont {Johnson}}, \bibinfo {author} {\bibfnamefont
  {J.}~\bibnamefont {Nygård}}, \bibinfo {author} {\bibfnamefont {C.~M.}\
  \bibnamefont {Marcus}},\ and\ \bibinfo {author} {\bibfnamefont {T.~S.}\
  \bibnamefont {Jespersen}},\ }\href {https://doi.org/10.1038/nmat4176}
  {\bibfield  {journal} {\bibinfo  {journal} {Nature Materials}\ }\textbf
  {\bibinfo {volume} {14}},\ \bibinfo {pages} {400} (\bibinfo {year}
  {2015})}\BibitemShut {NoStop}%
\bibitem [{\citenamefont {Shabani}\ \emph {et~al.}(2016)\citenamefont
  {Shabani}, \citenamefont {Kjaergaard}, \citenamefont {Suominen},
  \citenamefont {Kim}, \citenamefont {Nichele}, \citenamefont {Pakrouski},
  \citenamefont {Stankevic}, \citenamefont {Lutchyn}, \citenamefont
  {Krogstrup}, \citenamefont {Feidenhans'l}, \citenamefont {Kraemer},
  \citenamefont {Nayak}, \citenamefont {Troyer}, \citenamefont {Marcus},\ and\
  \citenamefont {Palmstrøm}}]{shabani_two-dimensional_2016}%
  \BibitemOpen
  \bibfield  {author} {\bibinfo {author} {\bibfnamefont {J.}~\bibnamefont
  {Shabani}}, \bibinfo {author} {\bibfnamefont {M.}~\bibnamefont {Kjaergaard}},
  \bibinfo {author} {\bibfnamefont {H.~J.}\ \bibnamefont {Suominen}}, \bibinfo
  {author} {\bibfnamefont {Y.}~\bibnamefont {Kim}}, \bibinfo {author}
  {\bibfnamefont {F.}~\bibnamefont {Nichele}}, \bibinfo {author} {\bibfnamefont
  {K.}~\bibnamefont {Pakrouski}}, \bibinfo {author} {\bibfnamefont
  {T.}~\bibnamefont {Stankevic}}, \bibinfo {author} {\bibfnamefont {R.~M.}\
  \bibnamefont {Lutchyn}}, \bibinfo {author} {\bibfnamefont {P.}~\bibnamefont
  {Krogstrup}}, \bibinfo {author} {\bibfnamefont {R.}~\bibnamefont
  {Feidenhans'l}}, \bibinfo {author} {\bibfnamefont {S.}~\bibnamefont
  {Kraemer}}, \bibinfo {author} {\bibfnamefont {C.}~\bibnamefont {Nayak}},
  \bibinfo {author} {\bibfnamefont {M.}~\bibnamefont {Troyer}}, \bibinfo
  {author} {\bibfnamefont {C.~M.}\ \bibnamefont {Marcus}},\ and\ \bibinfo
  {author} {\bibfnamefont {C.~J.}\ \bibnamefont {Palmstrøm}},\ }\href
  {https://doi.org/10.1103/PhysRevB.93.155402} {\bibfield  {journal} {\bibinfo
  {journal} {Physical Review B}\ }\textbf {\bibinfo {volume} {93}},\ \bibinfo
  {pages} {155402} (\bibinfo {year} {2016})}\BibitemShut {NoStop}%
\bibitem [{\citenamefont {Bozkurt}\ and\ \citenamefont
  {Fatemi}(2023)}]{bozkurt_josephson_2023}%
  \BibitemOpen
  \bibfield  {author} {\bibinfo {author} {\bibfnamefont {A.~M.}\ \bibnamefont
  {Bozkurt}}\ and\ \bibinfo {author} {\bibfnamefont {V.}~\bibnamefont
  {Fatemi}},\ }in\ \href {https://doi.org/10.1117/12.2678477} {\emph {\bibinfo
  {booktitle} {Spintronics {XVI}}}},\ Vol.\ \bibinfo {volume} {12656}\
  (\bibinfo  {publisher} {SPIE},\ \bibinfo {year} {2023})\ pp.\ \bibinfo
  {pages} {35--43}\BibitemShut {NoStop}%
\bibitem [{\citenamefont {Aumentado}\ \emph {et~al.}(2004)\citenamefont
  {Aumentado}, \citenamefont {Keller}, \citenamefont {Martinis},\ and\
  \citenamefont {Devoret}}]{aumentado2004Nonequilibrium}%
  \BibitemOpen
  \bibfield  {author} {\bibinfo {author} {\bibfnamefont {J.}~\bibnamefont
  {Aumentado}}, \bibinfo {author} {\bibfnamefont {M.~W.}\ \bibnamefont
  {Keller}}, \bibinfo {author} {\bibfnamefont {J.~M.}\ \bibnamefont
  {Martinis}},\ and\ \bibinfo {author} {\bibfnamefont {M.~H.}\ \bibnamefont
  {Devoret}},\ }\href {https://doi.org/10.1103/PhysRevLett.92.066802}
  {\bibfield  {journal} {\bibinfo  {journal} {Phys. Rev. Lett.}\ }\textbf
  {\bibinfo {volume} {92}},\ \bibinfo {pages} {066802} (\bibinfo {year}
  {2004})}\BibitemShut {NoStop}%
\bibitem [{\citenamefont {{Mart{\'i}n-Rodero}}\ and\ \citenamefont
  {Levy~Yeyati}(2011)}]{martin-rodero2011Josephson}%
  \BibitemOpen
  \bibfield  {author} {\bibinfo {author} {\bibfnamefont {A.}~\bibnamefont
  {{Mart{\'i}n-Rodero}}}\ and\ \bibinfo {author} {\bibfnamefont
  {A.}~\bibnamefont {Levy~Yeyati}},\ }\href
  {https://doi.org/10.1080/00018732.2011.624266} {\bibfield  {journal}
  {\bibinfo  {journal} {Advances in Physics}\ }\textbf {\bibinfo {volume}
  {60}},\ \bibinfo {pages} {899} (\bibinfo {year} {2011})}\BibitemShut
  {NoStop}%
\bibitem [{\citenamefont {Kadlecová}\ \emph {et~al.}(2019)\citenamefont
  {Kadlecová}, \citenamefont {Žonda}, \citenamefont {Pokorný},\ and\
  \citenamefont {Novotný}}]{kadlecova_practical_2019}%
  \BibitemOpen
  \bibfield  {author} {\bibinfo {author} {\bibfnamefont {A.}~\bibnamefont
  {Kadlecová}}, \bibinfo {author} {\bibfnamefont {M.}~\bibnamefont {Žonda}},
  \bibinfo {author} {\bibfnamefont {V.}~\bibnamefont {Pokorný}},\ and\
  \bibinfo {author} {\bibfnamefont {T.}~\bibnamefont {Novotný}},\ }\href
  {https://doi.org/10.1103/PhysRevApplied.11.044094} {\bibfield  {journal}
  {\bibinfo  {journal} {Physical Review Applied}\ }\textbf {\bibinfo {volume}
  {11}},\ \bibinfo {pages} {044094} (\bibinfo {year} {2019})}\BibitemShut
  {NoStop}%
\bibitem [{\citenamefont {Bargerbos}\ \emph {et~al.}(2022)\citenamefont
  {Bargerbos}, \citenamefont {Pita-Vidal}, \citenamefont {Žitko},
  \citenamefont {Ávila}, \citenamefont {Splitthoff}, \citenamefont
  {Grünhaupt}, \citenamefont {Wesdorp}, \citenamefont {Andersen},
  \citenamefont {Liu}, \citenamefont {Kouwenhoven}, \citenamefont {Aguado},
  \citenamefont {Kou},\ and\ \citenamefont {van
  Heck}}]{bargerbos_singlet-doublet_2022}%
  \BibitemOpen
  \bibfield  {author} {\bibinfo {author} {\bibfnamefont {A.}~\bibnamefont
  {Bargerbos}}, \bibinfo {author} {\bibfnamefont {M.}~\bibnamefont
  {Pita-Vidal}}, \bibinfo {author} {\bibfnamefont {R.}~\bibnamefont {Žitko}},
  \bibinfo {author} {\bibfnamefont {J.}~\bibnamefont {Ávila}}, \bibinfo
  {author} {\bibfnamefont {L.~J.}\ \bibnamefont {Splitthoff}}, \bibinfo
  {author} {\bibfnamefont {L.}~\bibnamefont {Grünhaupt}}, \bibinfo {author}
  {\bibfnamefont {J.~J.}\ \bibnamefont {Wesdorp}}, \bibinfo {author}
  {\bibfnamefont {C.~K.}\ \bibnamefont {Andersen}}, \bibinfo {author}
  {\bibfnamefont {Y.}~\bibnamefont {Liu}}, \bibinfo {author} {\bibfnamefont
  {L.~P.}\ \bibnamefont {Kouwenhoven}}, \bibinfo {author} {\bibfnamefont
  {R.}~\bibnamefont {Aguado}}, \bibinfo {author} {\bibfnamefont
  {A.}~\bibnamefont {Kou}},\ and\ \bibinfo {author} {\bibfnamefont
  {B.}~\bibnamefont {van Heck}},\ }\href
  {https://doi.org/10.1103/PRXQuantum.3.030311} {\bibfield  {journal} {\bibinfo
   {journal} {PRX Quantum}\ }\textbf {\bibinfo {volume} {3}},\ \bibinfo {pages}
  {030311} (\bibinfo {year} {2022})}\BibitemShut {NoStop}%
\bibitem [{\citenamefont {Glazman}\ and\ \citenamefont
  {Catelani}(2021)}]{glazman_bogoliubov_2021}%
  \BibitemOpen
  \bibfield  {author} {\bibinfo {author} {\bibfnamefont {L.}~\bibnamefont
  {Glazman}}\ and\ \bibinfo {author} {\bibfnamefont {G.}~\bibnamefont
  {Catelani}},\ }\bibfield  {journal} {\bibinfo  {journal} {SciPost Physics
  Lecture Notes}\ }\href {https://doi.org/10.21468/SciPostPhysLectNotes.31}
  {10.21468/SciPostPhysLectNotes.31} (\bibinfo {year} {2021})\BibitemShut
  {NoStop}%
\end{thebibliography}%

\end{document}